\documentclass[printer]{aa}

\usepackage{natbib}
\usepackage{amsmath}
\usepackage{graphicx}
\usepackage{txfonts}
\usepackage{color}
\usepackage{threeparttable}

\begin{document}

\title{Probing spacetime around Sagittarius~A* using modeled VLBI closure phases}
\titlerunning{Closure phases}
\authorrunning{Fraga-Encinas, R. et al. }
\author{R. Fraga-Encinas\inst{1},
                M. Mo{\'s}cibrodzka\inst{1},
                C. Brinkerink\inst{1},
                H. Falcke\inst{1}}
\institute{$^1$Department of Astrophysics / IMAPP, Radboud University, 
P.O. Box 9010, 6500 GL Nijmegen, The Netherlands\\
\email{r.fraga@astro.ru.nl}}
\date{Accepted 2015 December 27. Received 2015 October 10.}

\abstract
    {The emission region and black hole shadow of Sagittarius A*, the
      supermassive black hole at the Galactic Center, 
      can be probed with millimeter Very Long Baseline Interferometry.}                                                            
    {Our goal is to probe the geometry of the emitting plasma around Sgr
      A* by using modeled mm-VLBI closure phase calculations at 1.3 mm 
      and to constrain the observer's inclination angle and position angle of the black hole spin axis.}
    {We have simulated images for three different models of the emission of
      Sgr A*: an orbiting spot, a disk model, and a jet model. 
      The orbiting spot model was used as a test case scenario,
      while the disk and jet models are physically driven scenarios 
      based on standard three-dimensional general relativistic 
      magnetohydrodynamic simulations of hot accretion flows. 
      Our results are compared to currently available closure phase observational limits.}
      {Our results indicate that more models with closer to edge-on
      viewing angles are consistent with observational limits. In general,
      jet and disk geometries can reproduce similar closure phases for
      different sets of viewing and position angles. Consequently, the favored
      black hole spin orientation and its magnitude are strongly model dependent. 
      }
      {We find that both the jet and the disk models can explain current VLBI
      limits. We conclude that new observations at 1.3 mm and possibly at longer
      wavelengths including other triangles of VLBI baselines are necessary to
      interpret Sgr A* emission and the putative black hole spin parameters.}


   \maketitle

\section{Introduction}

The compact radio source at the center of the Milky Way, Sagittarius A*
(Sgr~A*), is the strongest candidate for a supermassive black hole (SMBH)
(\citealt{ghez:2000}, \citealt{genzel:2010}; \citealt{falcke:2013}). Stellar orbits indicate a
central object mass of about four million solar masses concentrated in a very
small volume (\citealt{eisenhauer:2005}; \citealt{ghez:2008}; \citealt{gillessen:2009}). Sgr~A* is the closest known SMBH to us,
and it is an excellent candidate for imaging the shadow of an event
horizon \citep{falcke:2000} and for testing general relativity (\citealt{broderick:2014}; \citealt{ricarte:2015}). Its angular diameter on the sky is approximately 50
$\mu as,$ which can be resolved by the currently constructed Event Horizon
Telescope (EHT) \citep{doeleman:2009_whitepaper}.

The EHT is a global Very Long Baseline Interferometry
(VLBI) experiment aimed at imaging Sgr~A* (and the core of M87) at $\lambda$=1.3 mm and
0.87 mm (230 GHz and 345 GHz, respectively). Here, the positive spectral index at the lower frequencies changes to a negative spectral index at higher frequencies, which is suggestive of the plasma becoming optically thin (\citealt{bower:2015}) and is crucial to take an unobscured picture of the SMBH shadow and plasma around it. Additionally, at millimeter
wavelengths the interstellar medium scattering screen smearing effects become
negligible and so the EHT promises to take sharp images of Sgr~A*. Various
theoretical models have been developed and used to construct the expected source
appearance at EHT wavelengths (e.g., \citealt{broderick:2006},
\citealt{doeleman:2009}, \citealt{moscibrodzka:2009}, \citealt{dexter:2010,dexter:2011}, \citealt{moscibrodzka:2012}, 
\citealt{dolence:2012}, \citealt{roman:2012}, \citealt{kamruddin:2013}, \citealt{moscibrodzka:2014} and references therein).

In addition to image reconstruction, the geometry of Sgr~A* can be constrained by 
using closure phases. This method
is particularly useful if a small number of VLBI stations is available (as is the case
at present) and the \textit{uv}-coverage in the Fourier domain is too sparse to
directly and accurately reconstruct the image from the VLBI visibilities.

The goal of this work is to investigate the geometry of the plasma near the SMBH
using non-imaging VLBI techniques. Based on various
theoretical models of plasma near the SMBH, we here present the expected closure
phase variations for the current triangles of EHT stations and compare them with observational
limits. The closure phases are
predicted for 1.3 mm VLBI observations for the triangle formed by baselines including existing EHT
stations (CARMA, Hawaii, SMTO) at which Sgr~A* has already been detected and
observational constraints exist. For one of the models, which
is used as a  simple test case, we also present closure phases for  triangles including European (IRAM PV, IRAM PdB) and American stations (CARMA, SMTO, LMT, Hawaii, Chile). Information on all of the stations is summarized in Table~\ref{tab:SEFD}.

Determining the intrinsic size and shape of Sgr~A* are challenging problems.  Measurements of the source morphology are highly affected by calibration uncertainties. These uncertainties arise from the variability of antenna gains and the atmospheric opacity that is due to the low antenna elevations that are needed to
observe Sgr~A* from the Northern hemisphere. Because closure quantities such as closure amplitude and closure phase are independent of antenna-based amplitude errors and phase shifts due to atmospheric turbulence, they are extremely handy non-imaging tools used to determine the structure of a source (\citealt{rogers:1974}, \citealt{cotton:1979}, \citealt{bower:2004}). 

These closure quantities have previously been used to constrain the structure of Sgr~A* at mm wavelengths. VLBI observations at 3.5 mm resulted in some estimates for the size of Sgr~A*; the data were modeled assuming a circular Gaussian brightness distribution. The data were also consistent with an elliptical Gaussian structure (with an elongation in the north-south direction) due to interstellar scattering of a point source (\citealt{doeleman:2001}). Additional VLBI observations at 3.5 mm using the array formed by the Effelsberg telescope, the IRAM Pico Veleta telescope, and the IRAM Plateau de Bure interferometer yielded a closure phase measurement of $0\degr\pm 10\degr$ (\citealt{krichbaum:2006}). 

Simulated images of jets at 7 mm (43 GHz) based on models consistent with the spectrum
of Sgr~A* have also been constructed and compared to 7 mm VLBI observations
through closure quantities (\citealt{markoff:2007}). These jet models with a bipolar
structure, high-inclination with respect to the line of sight of $\theta
\gtrapprox 75\degr$, and position angle in the sky of $105\degr$ east of north
produced results that are comparable to Gaussian models from previously done work. Furthermore, by using 7 mm VLBI observations and closure quantities, limits have been placed on the size versus wavelength relation (\citealt{bower:2004}, \citealt{bower:2006}), where the intrinsic size of Sgr~A* decreases with observing wavelength following a power law with index of $\sim 1.6$. The implications are that the source is optically thick and presents a photosphere that changes in size depending on the observing wavelength. 

1.3 mm VLBI closure phases have also been used to constrain the structure of
the accretion flow of Sgr~A*. Radiatively inefficient accretion flow models
(RIAFs, as described in \citealt{broderick:2011_models}) consistent with
observed 1.3 mm VLBI visibility amplitudes produced closure phases of $0\degr\pm 30\degr$ with a predicted most probable value of ~ $0\degr\pm 13\degr$ (\citealt{broderick:2011_closure}) on the triangle of baselines formed by CARMA-SMTO-JCMT. VLBI observations at 1.3 mm on the same triangle of stations reported, for the first time at this wavelength, a closure phase measurement of $0\degr\pm 40\degr$ (\citealt{fish:2011}). An increase in the flux density during one of the nights was detected, which is an indicator of time-variability of Sgr~A* at Schwarzschild radius scales. 

Since the structure of Sgr~A* still remains elusive, the work in this paper
explores three different emission models for Sgr~A* at 1.3 mm: an orbiting hot
spot, a disk, and a jet model. The hot spot model is presented as a test case to compare our results with the work done by \citealt{doeleman:2009},
where a hot spot embedded in an accretion disk could be a plausible
explanation for the flares observed in Sgr~A* in the NIR
(\citealt{genzel:2003}, \citealt{belanger:2006}, \citealt{ghez:2004}) and in
the X-ray band (\citealt{baganoff:2001}). For the disk and jet models, the three-dimensional general relativistic magnetohydrodynamic (3D GRMHD) 
simulations are used to model the emission of magnetized plasma accreting onto
a spinning SMBH (\citealt{moscibrodzka:2013,moscibrodzka:2014} and references
therein). These 3D GRMHD models are investigated because they can provide a more physically accurate description of the accretion flow than an orbiting hot spot model (\citealt{broderick:2006}) or an adiabatically expanding blob model (\citealt{yusef-zadeh:2009}). These relativistic magnetohydrodynamic (MHD) models are well suited for modeling millimeter emission that comes from the areas closest to the black hole, and therefore regions where relativistic effects are strong. In previous work, time-dependent images of millimeter
synchrotron emission from a 3D GRMHD accretion disk model
  were fitted to mm-VLBI data (\citealt{dexter:2009}). The models of
the accretion flow were explored to estimate values for the inclination of the
accretion disk with respect to the observer, position angle in the sky,
accretion rate, and electron temperature of the accretion flow
(\citealt{dexter:2010}). In the work presented here, we introduce
for the first time in addition to a relativistic disk model a jet model based on 3D GRMHD simulations. 
Our goal is to use the modeled and observed visibility closure phases to
place constraints on the possible structures of Sgr~A* and putative black hole
spin orientation in space.

The paper is organized as follows. In Sect.~\ref{closure_phases} we recall
basic definitions and properties of VLBI visibility amplitudes, visibility
phases, and closure phases. In Sect.~\ref{emission_models} we present
theoretical, general relativistic models of Sgr~A*.
In Sect.~\ref{data} we describe the method for simulating the closure phase
observations and the parameters we used to reconstruct the observational noise.
We present the model appearances and the resulting closure phase evolution for 
selected VLBI triangles in Sect.~\ref{results}. We discuss the results in
Sect.~\ref{discussion} and conclude in Sect.~\ref{conclusions}.

\section{Closure phases}~\label{closure_phases}
\subsection{Closure phase calculation}
The visibility function is a Fourier transform of the 
intensity distribution on the sky and is given by
\begin{equation}\label{eq:visibility}
V(u,v)= \iint I(x,y) ~e^{-2 \pi i(ux+vy)} ~dx dy 
,\end{equation}
where $I(x,y)$ is the intensity distribution at a given set of coordinates
$x,y$ on the sky (the $x,y$ angular coordinates are left-handed, i.e., $x$ and $y$ 
are positive in east and north directions on the sky, respectively), 
and $u,v$ are the projected (also left-handed) baseline lengths.
The $V(u,v)$ is by definition a complex function, so it has an amplitude, $A$,
and a phase, $\phi$,
\begin{equation}\label{eq:visibility_complex}
V(u,v)= A ~e^{-i \phi}
.\end{equation}

The sum of visibility phases around a closed loop is known as
the closure phase, first introduced by \citet{jennison:1958}. The closure
phase of a triangle formed by the baselines between stations \textit{i, j},
and \textit{k} is the sum of the visibility phases on each baseline:
\begin{equation}
\Phi_{ijk}\equiv\phi_{ij}+\phi_{jk}+\phi_{ki} ~.
\end{equation}
Hence, the closure phase is the sum of the arguments of the complex
visibilities on the baselines forming a triangle of stations:
\begin{equation}\label{eq:closure_phase}
\Phi_{ijk}=\arg[V(u_{ij},v_{ij})]+\arg[V(u_{jk},v_{jk})]+\arg[V(u_{ki},v_{ki})] ~.
\end{equation}

The closure phase is a good observable because it is unaffected by phase
errors introduced by individual stations due to atmospheric turbulence or instrumental
instabilities. Consider three stations \textit{i,j,k}. A blob of air with high moisture content over station \textit{j} will introduce a phase delay above this station, so that the fringes detected by the baselines formed by \textit{i} and \textit{j} will be shifted in phase. However, because there will also be an equal but opposite phase shift in the fringes detected by stations \textit{j} and \textit{k}, then the overall sum of phases over the triangle of stations will be insensitive to the phase delay introduced by station \textit{j}.
For point sources, Gaussian distributions (symmetric and elliptical), and annular distributions 
the closure phase is always equal to 0$^{\circ}$ or 180$^{\circ}$. Non-zero (non-180$\degr$) values of closure phase indicate
asymmetries or skewness in the source structure (\citealt{monnier:2007}). For Sgr~A*, on the triangle formed by the CARMA-SMTO-JCMT baselines, a closure phase of 0$^{\circ}$ $\pm$ 40$^{\circ}$
has been measured at 1.3 mm \citep{fish:2011}.

\subsection{Closure phase errors}~\label{closure_phase_errors}
The errors on the closure phases are dependent on the signal-to-noise-ratio
(S/N, $s_{ij}$) of the complex visibilities on individual baselines
(\citealt{broderick:2011_closure}),
\begin{equation}\label{eq:snr_visibilities}
s_{ij}\equiv |V_{ij}| \sqrt{\frac{2B\tau}{\mathrm{SEFD}_{i}~\mathrm{SEFD}_{j}}} ~ ,
\end{equation}
where $|V_{ij}|$ is the complex visibility amplitude, $B$ is the bandwidth,
$\tau$ is the coherence time of the atmosphere, and $\mathrm{SEFD}_{i}$ is the
system equivalent flux density for a given station. Values for the
coherence time, which depends on the observing wavelength, at these sites can range from a few seconds to $\tau$ $\sim$ 20 s under good weather conditions (\citealt{doeleman:2002}). For our work, we chose a value of $\tau$=20 s. Since, as we previously
mentioned, the closure phases are independent of phase delays introduced by
atmospheric instabilities, they can be averaged over timescales longer than
$\tau$. Nevertheless, we have to take into account some limiting factors such
as the timescales on which the structure of Sgr~A* and the orientation of the
baselines do not change significantly. When Sgr~A* is not in a quiescent state,
that is, flaring, these timescales could be shorter. The bandwidth is chosen
to be one of the observing bandwidths for the EHT with $B$=4 GHz of 
total on-sky bandwidth with 2 GHz per polarization. 
The values for the $\mathrm{SEFD}$ of given observatories adopted from \citealt{doeleman:2009} are listed in
Table~\ref{tab:SEFD}.

\begin{table*}
  \centering
  \tiny
  \caption{Estimated system equivalent flux densities (SEFD) at 1.3mm. 
    The diameter is the effective aperture when the given
    number of antennas are phased together. The expected SEFD
    values for observations of Sgr~A* include typical weather conditions and
    opacities.}
  \begin{threeparttable}
  \begin{tabular}{l l c c c}
    \hline\hline
    Facility Acronym   & Facility Name                                                        & Antennas     & Diameter   & SEFD  \\
                       &                                                                        &                    &  (m)             & (Jy)  \\
    \hline
    Hawaii     & 3 facilities are phased: Caltech Submillimeter Observatory (CSO),               &            &            &       \\
               & James Clerk Maxwell Telescope (JCMT) and Submillimeter Array (SMA)              & 8          &  23        & 4,900  \\
    CARMA          & Combined Array for Research in Millimeter Wave Astronomy                        & 8          &  27        & 6,500  \\
    SMTO       & Arizona Radio Observatory Submillimeter Telescope                               & 1          &  10        & 11,900  \\     
    LMT        & Large Millimeter Telescope                                                      & 1          &  32        & 10,000\tnote{a} \\
    APEX       & Atacama Pathfinder Experiment                                                   & 1          &  12        & 6,500   \\
    ALMA       & Atacama Large Millimeter Array (ALMA)                                           & 10         &  38        & 500     \\
    IRAM-PV    & 30-m Institut de Radioastronomie Millim{\'e}trique telescope on Pico Veleta     & 1          &  30        & 2,900   \\ 
    IRAM-PdB   & Institut de Radioastronomie Millim{\'e}trique interferometer on Plateau de Bure & 6          &  15        & 1,600   \\ [1ex]
    \hline
  \end{tabular}
  \begin{tablenotes}   
  \item[a] Upgrades to the dish and receiver will lower the SEFD of the LMT by a factor of $\sim$ 10.
  \end{tablenotes}
  \end{threeparttable}   
  \label{tab:SEFD}
\end{table*}

The noise estimates can be calculated as shown in \citet{rogers:1995}:
\begin{equation}\label{eq:closure_phase_error}
\sigma_{\Phi_{ijk}} = \frac{\sqrt{4+s_{ij}^2 s_{jk}^2+s_{ij}^2
    s_{ki}^2+s_{jk}^2 s_{ki}^2+2(s_{ij}^2+s_{jk}^2+s_{ki}^2)}}{(s_{ij} s_{jk}
  s_{ki})(T/\tau)^{1/2}} \, [rad]
,\end{equation}
where $T$ is the integration time. It is important to note that $\sigma_{\Phi_{ijk}}$ is model dependent because
its value depends on the S/N of the complex visibilities on individual
baselines. Equation~\ref{eq:closure_phase_error} 
for high S/N ($s_{ij},s_{jk},s_{ki}\gg1$) converges to
Eq.11 in \citet{broderick:2011_closure}\footnote{Note the typo in power in
Eq.11 in \citet{broderick:2011_closure}.}.

\section{Theoretical emission models}\label{emission_models}

The theoretical emission images are produced using the general relativistic
ray-tracing method used in \citet{moscibrodzka:2009,moscibrodzka:2014}. These
numerical ray-tracing calculations simulate the radiative transfer equations for
synchrotron radiation from a thermal distribution of electrons near a black hole.

To simulate the effects of source smearing by free electrons in the
Galaxy, we convolved all the theoretical images of the source
with the elliptical scattering Gaussian function. The Gaussian parameters were
adopted from radio observations of the source at long wavelengths at
  which the source is completely dominated by the scattering \citep{bower:2004,bower:2006}.
We note that the adopted scattering screen is described by a 
function for which the complex visibility only has a non-zero real part.  
 
For each model (described in detail in the next subsections), we created frames showing
the observed image of the model every 10 seconds over the 12-hour observation
interval (the source is visible at various VLBI stations at different times).
For each model, we created frames with a resolution of $128 \times 128$~ pixels
of the Milky Way central $40 \times 40 R_g$. The $R_g=GM/c^2$ is the
gravitational radius of the SMBH and for Sgr~A* $R_g=6.6 \times 10^{11} {\rm
  cm}$. Therefore our modeled field of view is about $200 \times 200$ $\mu as$
on the sky. The visibility amplitude and phase at each moment in time were
computed from the complex visibility function produced by a 2D Fourier
transformation of the corresponding theoretical image
(Eq.~\ref{eq:visibility}).

We considered three emission models.  The first, an orbiting spot model, is
a simplified model of a quiescent accretion flow with a variable component to
simulate a flaring event, and the computations were carried out
for reference only. The other two, the disk and jet models, are realistic
models of emission from an accreting black hole derived from
3D GRMHD simulations
(\citealt{moscibrodzka:2013,moscibrodzka:2014} and references therein).

\subsection{Images of the orbiting spot}~\label{spot}

Our orbiting spot model is similar to the spot model studied in
\citet{doeleman:2009} (and references therein). In this model, the
background radiation (or a 'quiescent', stationary emission) 
is produced by a RIAF (see, e.g., \citealt{broderick:2011_models}) onto a spinning black hole
(hereafter $a_*$ is the dimensionless spin of the SMBH). In our RIAF model,
the plasma number density, the electron temperature (always defined in electron
rest mass units, i.e., $\Theta_e=kT_e/m_ec^2$), and the magnetic field strength 
are constant in time and have the following radial distributions:
\begin{equation} \label{eq:n_density}
n_{e}= n_{e}^0 (\frac{r}{R_g})^{-1.1} exp(z^2/2r^2),
\end{equation}
\begin{equation}\label{eq:e_temp}
\Theta_e=\Theta_e^0 (\frac{r}{R_g})^{-0.84},
\end{equation}
and 
\begin{equation}\label{eq:b_field}
\frac{B^2}{8\pi}=\beta^{-1} n_{e} \frac{ m_pc^2 R_g}{6r}
,\end{equation}
where $n_e^0$, $\Theta_e^0$, and $\beta$ (a parameter describing the ratio of gas to
magnetic field pressure, $\beta=P_{gas}/P_{mag}$) are the model free parameters.
The RIAF rotates around the central object with
a Keplerian angular velocity:
\begin{equation}
\Omega_{K} (r,a_*)= \frac{1}{(r/R_g)^{3/2}+a_*}
\end{equation}
and has a zero radial velocity. 
The model variable component, the orbiting spot, as in \citet{doeleman:2009}, 
is described by a Gaussian shape of size $R_{spot}=0.75 R_s$ 
and it is orbiting at a Keplerian orbit at the equatorial plane of the BH at a radius
$r_{spot}$ (model parameter).
The density inside of a spot is enhanced: 
\begin{equation}
n_{e,spot}=n^0_{e,spot} \exp{\left(-\frac{|\vec{x}-\vec{x}_{spot}(t)|^2}{2 R_{spot}^2}\right)}
,\end{equation}
where $|\vec{x}-\vec{x}_{spot}(t)|$ is the varying
distance between the photon geodesics and the spot center 
to account for a delay between the observer time and the 
current coordinate time at the position of the spot (for
more details on the spot model see, e.g., \citealt{schnittman:2006}).

To check the consistency of our computations with the
results presented in \citet{doeleman:2009}, we adopted free parameters of 
the model to reconstruct models similar to runs A230, B230, C230, and D230 
shown in their work. Our model parameters are summarized in Table~\ref{tab:model_params}.

\begin{table*}
\centering
\tiny
\caption{ Summary of parameters for four orbiting spot models (A, B, C,
  and D). Parameters include the black hole spin ($a_*$), the spot orbital period
  ($P$), observer's inclination angle ($i$) which is the angle between the observer's line of sight and the black hole spin axis, position angle of the black
  hole spin axis on the sky ($PA$,
  note that \citealt{doeleman:2009} defined $PA$ as a disk major axis
  position angle, i.e., their $PA$ is offset by $+90\degr$ from our values), observing frequency ($\nu$), ADAF flux ($Disk$), minimum and maximum fluxes ($Min$ and $Max$), electron number density ($n_{e}^0$), electron temperature ($\Theta_{e}^0$), gas-to-magnetic-field-pressure parameter ($\beta$), the spot orbital radius
  ($r_{spot}$), and the spot electron density ($n_{e,spot}^0$).}\label{tab:model_params}
\begin{tabular}{c c c c c c c c c c c c c c c}
\hline\hline
ID & $a_*$ & P[min] & i[$\degr$] & PA[$\degr$] & $\nu$ [GHz] &
Disk [Jy] & Min [Jy] & Max[Jy] & $n_{e}^0 [{\rm cm^{-3}}]$ & $\Theta_{e}^0$ & $\beta$ &
 $r_{spot} [R_{\rm g}]$&$n_{e,spot}^0 [{\rm cm^{-3}}]$ \\
\hline
A & 0   & 27  & 30 & 0 & 230 & 2.1 & 3.02 & 4.46 & $4.3\times 10^6$ & 80 & 10 &5.5234 &$1.5\times 10^7$\\
B & 0   & 27  & 60 & 0 & 230 & 2.6 & 2.80 & 4.44 & $4.0\times 10^6$ & 80 & 10 &5.5234 &$4.5\times 10^6$\\
C & 0   & 27  & 60 & -90  & 230 & 2.6 & 2.80 & 4.44 & $4.0\times 10^6$ & 80 & 10 &5.5234 &$4.5\times 10^6$\\
D & 0.9 & 27  & 60 & 0 & 230 & 2.2 & 2.37 & 3.85 & $3.0\times 10^6$ & 80 & 10 &5.2650 &$4.0\times 10^6$\\ 
\hline
\end{tabular}
\end{table*}

\subsection{Images of disk and jet based on GRMHD simulations}~\label{disk_jet_models}

The theoretical images of a disk and a jet were constructed by combining the
ray-tracing radiative transfer model with a 3D GRMHD
simulation of magnetized,
turbulent plasma accreting onto a spinning SMBH ($a_*\approx0.94$). In the
3D GRMHD simulations, the magnetized jet is naturally produced by preexisting, poloidal magnetic fields
and the spinning black hole. The plasma density and the magnetic field strength
used in the radiative transfer models are taken directly from the
simulations. The synchrotron emission that we observe is most probably
produced by electrons. The electron temperature, $T_e$, is not explicitly
computed in the current GRMHD simulations and so it has to be
parameterized. In this work, we used the following prescription for electron
temperatures:
\begin{equation}\label{eq:temp_ratio}
\frac{T_p}{T_e}= C_{disk} \frac{\beta^2}{1+\beta^2} + C_{jet} \frac{1}{1+\beta^2}
,\end{equation}
where $T_p$ is the temperature of protons (provided by the GRMHD simulations), and $T_p/T_e$ is the unknown proton-to-electron temperature ratio. 
We assumed that $T_p/T_e$ is a function of the
$\beta$ plasma parameter.
The coupling constants $C_{disk}$ and $C_{jet}$ describe the
proton-to-electron coupling in the weakly and strongly magnetized
plasma, respectively.
In the case of a weakly magnetized plasma $\beta \gg 1$ (e.g., inside of
a turbulent accretion disk), $T_p/T_e \to C_{disk}$. 
For a strongly magnetized plasma, $\beta\ll1$ (e.g., along the jet) $T_p/T_e \to C_{jet}$.

In particular, we assumed in our disk model that electrons are strongly
coupled to protons both in the disk and in the jet ($C_{disk}=1$,
$C_{jet}=1$). 
Since the plasma density is the highest in the equatorial plane of the accretion
disk, $C_{disk}=1$ will lead to an image with a bright disk.

In the jet model, the electrons are weakly coupled to protons in the accretion
disk ($C_{disk}=20$), but remain strongly coupled to protons in the jet ($C_{jet}=1$);
and synchrotron emission from the jet will overcome the disk
emission. It is worth mentioning that jets produced in the GRMHD simulations
have two components: a jet spine and a jet sheath. The spine of the jet is
strongly magnetized and has a low matter component, and therefore it does not
produce any detectable electromagnetic signal. The jet sheath is a thin layer
of outflowing gas surrounding the empty spine, it moves away from the BH
relatively slowly at the considered distances 
and is made of baryonic plasma that originates from the inner
parts of the accretion disk. As a result of the much higher matter content of the jet
sheath in comparison to the spine, any synchrotron emission produced by the jet
will be dominated by the sheath component.

Both models, disk and jet, were normalized to produce a similar total flux of
approximately 2 Jansky at $\lambda$=1.3 mm, in accordance with observations \citep{doeleman:2008}. The
renormalization was made by changing the mass accretion rate $\dot{M}$
(i.e., multiplying the matter densities in the entire model by a constant
scaling density factor).

As shown in \citet{moscibrodzka:2013} and in \citet{moscibrodzka:2014}, 
the different electron temperature prescriptions in regions around the
  SMBH defined as the disk and jet significantly change a single GRMHD model appearance
and the shape of its observed spectral energy distribution. Our current
prescription that defines electron temperatures in the jet and disk
has been slightly modified compared to that used 
in \citet{moscibrodzka:2014}. This was done to obtain smoother
images, that is, avoid sharp boundaries between the disk and jet zones 
(see, e.g., \citealt{moscibrodzka:2015}). 

Since our electron temperature prescriptions in the disk and jet models are
still robust, we time-averaged the images produced by time-dependent simulations
over the duration of a few hours. Hence, any closure phase variations, based on
the time-averaged disk and jet images, will be due to the 
VLBI baselines rotation due to Earth's rotation and
probing different \textit{uv}-values in the Fourier space. This
is not the case for the orbiting spot model, which is fully time-dependent. 

\section{Simulated data}~\label{data}

The closure phase observations were simulated using the 
images of the orbiting spot model, the time-averaged 
disk model, and the time-averaged jet
model. For a given $T=10$-second scan, the Fourier transform of the brightness
distribution was computed using Eq.~\ref{eq:visibility}, and the components of
the complex visibility function (amplitude and phase) were obtained for each
baseline. The S/N for individual baselines was calculated
using Eq.~\ref{eq:snr_visibilities} with a coherence time for the atmosphere
of $\tau$=20 s and the appropriate system equivalent flux densities given in
Table~\ref{tab:SEFD}. Values for the closure phase and its errors were
calculated using Eqs.~\ref{eq:closure_phase} and~\ref{eq:closure_phase_error}, respectively.

\section{Results}~\label{results}

\subsection{Theoretical closure phase evolution for the orbiting spot model}

\begin{figure*}
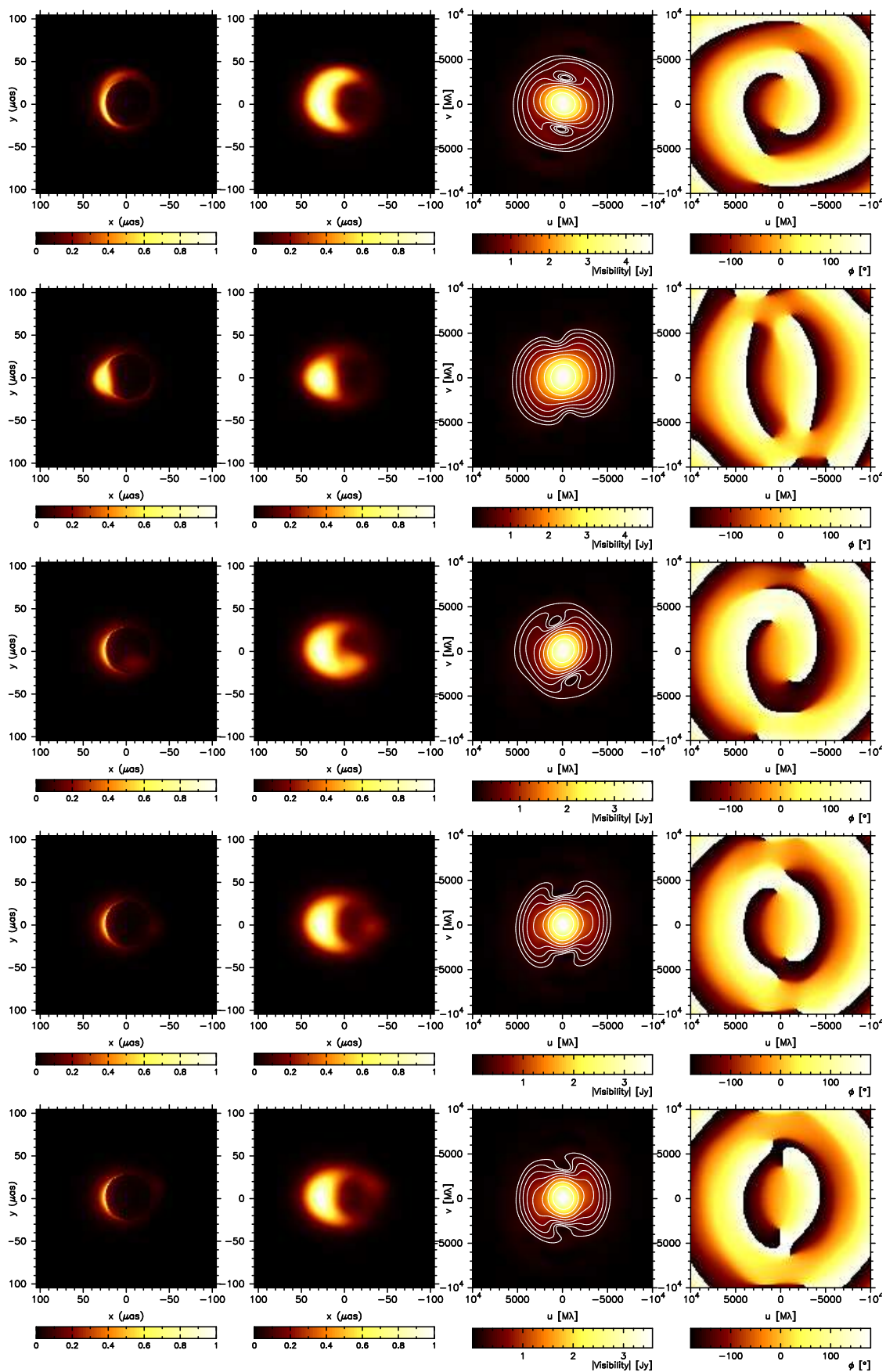

\centering
\includegraphics[width=0.25\textwidth,angle=-90]{modelB.phase0.0.new.ps}
\includegraphics[width=0.25\textwidth,angle=-90]{modelB.phase0.2.new.ps}
\includegraphics[width=0.25\textwidth,angle=-90]{modelB.phase0.4.new.ps}
\includegraphics[width=0.25\textwidth,angle=-90]{modelB.phase0.6.new.ps}
\includegraphics[width=0.25\textwidth,angle=-90]{modelB.phase0.8.new.ps}
\caption{Images of model B at $\lambda=$1.3mm (here $PA=0\degr$). Rows show orbital
  phases of the spot at t=0, 0.2, 0.4, 0.6, and 0.8 from top to bottom. Left to right panels
  show an image of the model, that image convolved with the scattering screen, the visibility amplitude, and the visibility phase of the scatter-broadened images. 
  The color intensity for the panels in the first two columns indicates the
  intensity of radiation, which has been normalized to unity. The visibility amplitude is in units of Jansky, and eighth contours are spaced by a factor of $\sqrt{2}$. The last column shows the corresponding map of the visibility phase. The range of uv values for Cols. 3 and 4 is the same.}\label{fig:spot_model}
\end{figure*}

As a test case, we studied the radiation and VLBI observables produced by the
orbiting spot model. Figure~\ref{fig:spot_model} shows
the orbiting spot model B (see Table~\ref{tab:model_params}) at various
moments in time. The panels from left to right show the
image of the model, the same image convolved with the scattering screen, the
visibility amplitude map with contours, and lastly
the phase of the visibility function. The rows
represent different orbital phases ranging from 0 to 0.8 from top to bottom.

The emission from RIAF that 
we see in the images of the model, where the peak brightness has been scaled to
1, has the shape of a crescent. The crescent is formed by 
light-bending effects because the
strong gravity dominates the SMBH surroundings as well as relativistic Doppler
beaming effects due to the Keplerian orbital motion of the gas. Relativistic
Doppler beaming of the approaching plasma causes the prograde orbiting spot
to become brighter only when moving toward the observer 
(i.e., see the second panel in
the first column), while the spot becomes dimmer as it recedes from the observer
(i.e., last two panels in the first column). In addition, the spot intensity is
not exactly Gaussian due to relativistic effects (time delays) of ray tracing. 
All the panels in the
second column show a much broader and brighter image of the crescent shape
because of the convolution with the scattering screen and because of the different amplitude scaling. Changes in the phase of
the visibility function are shown in the last column.

\begin{figure*}
\includegraphics[width=0.24\textwidth]{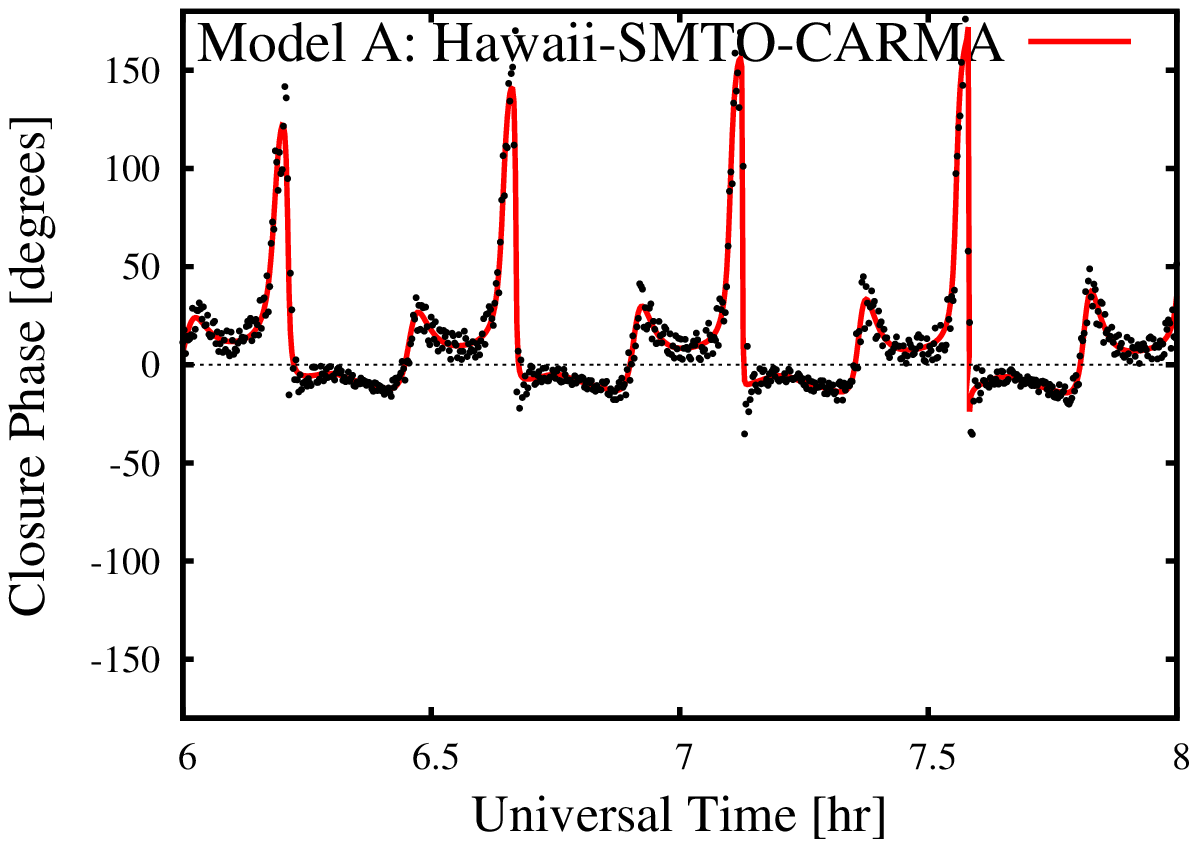}
\includegraphics[width=0.24\textwidth]{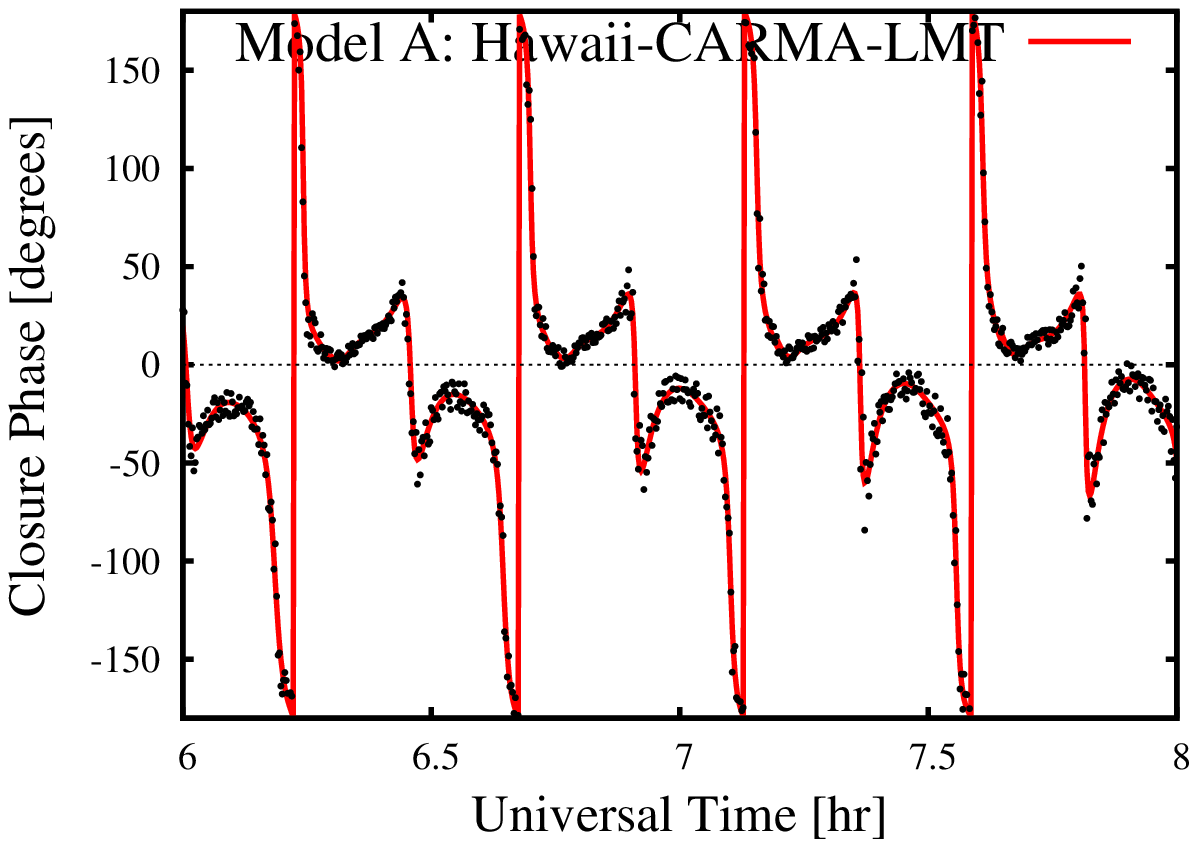}
\includegraphics[width=0.24\textwidth]{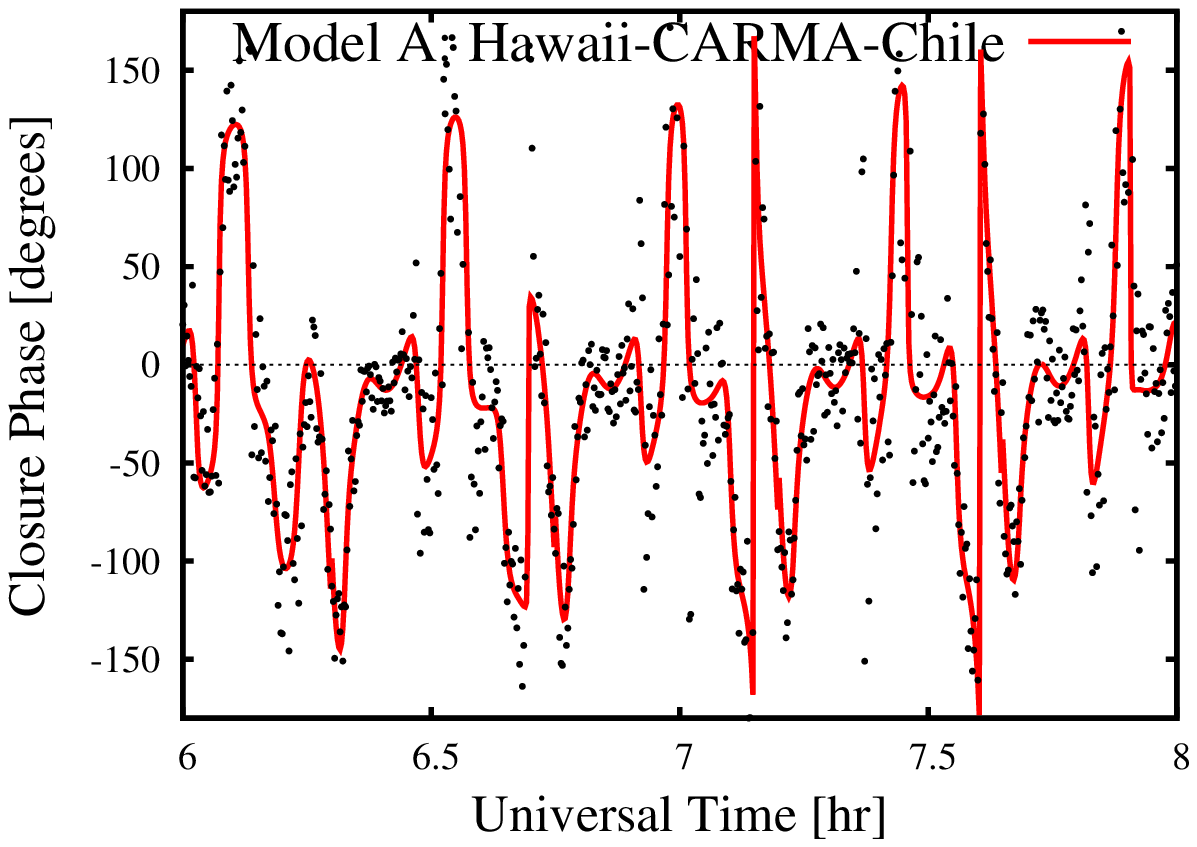}
\includegraphics[width=0.24\textwidth]{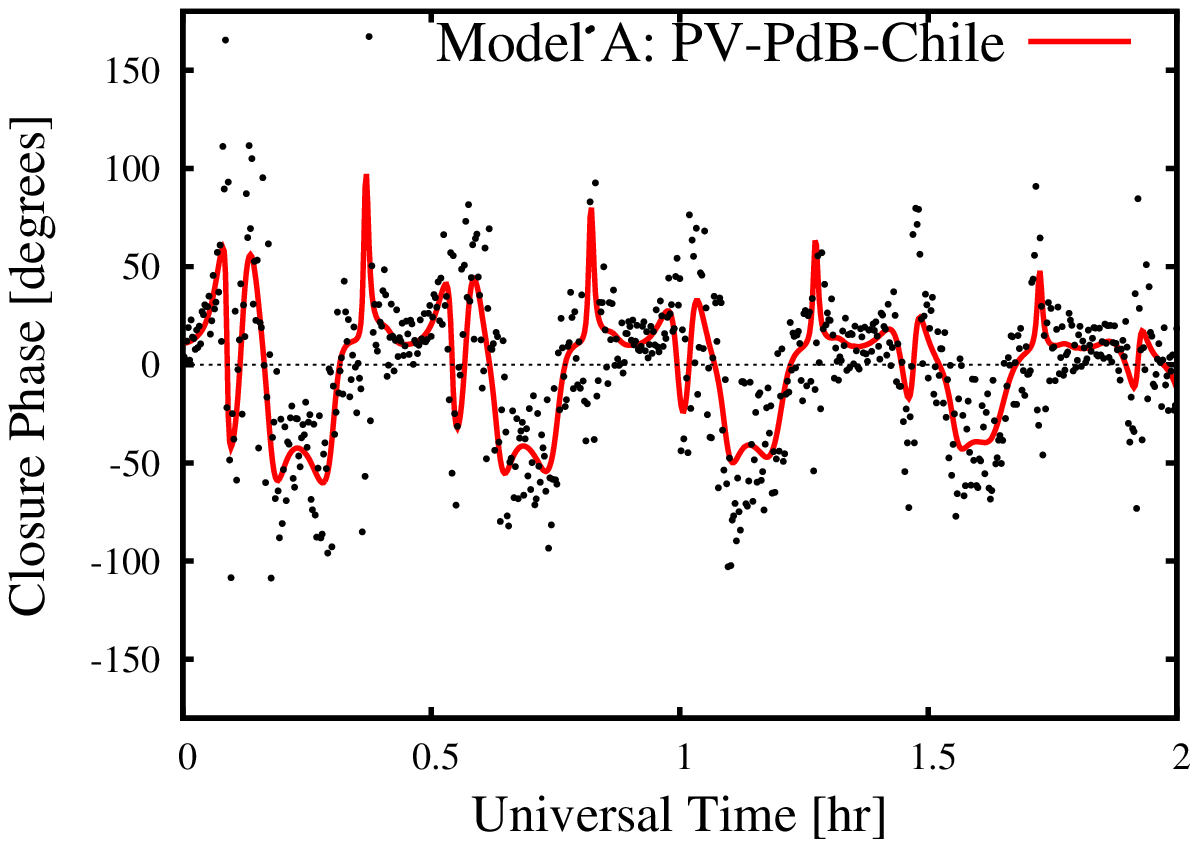}\\
\includegraphics[width=0.24\textwidth]{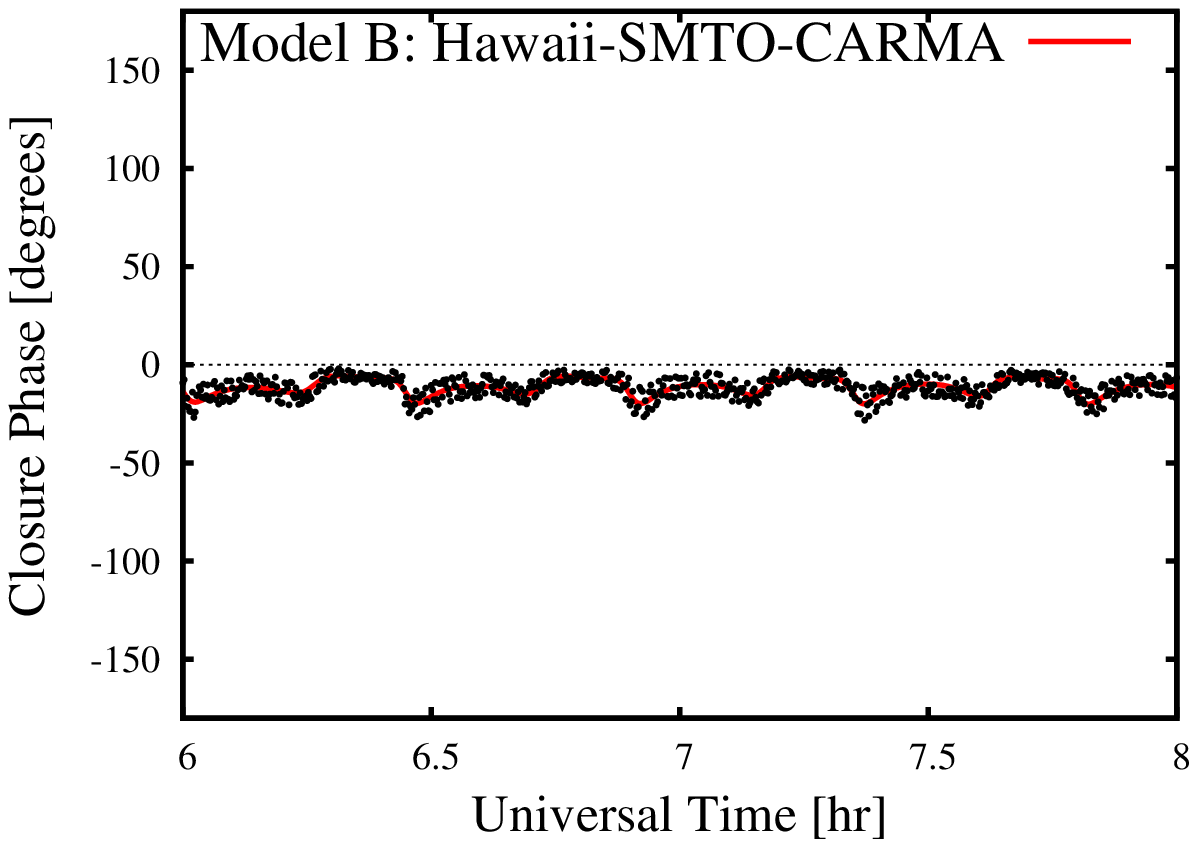}
\includegraphics[width=0.24\textwidth]{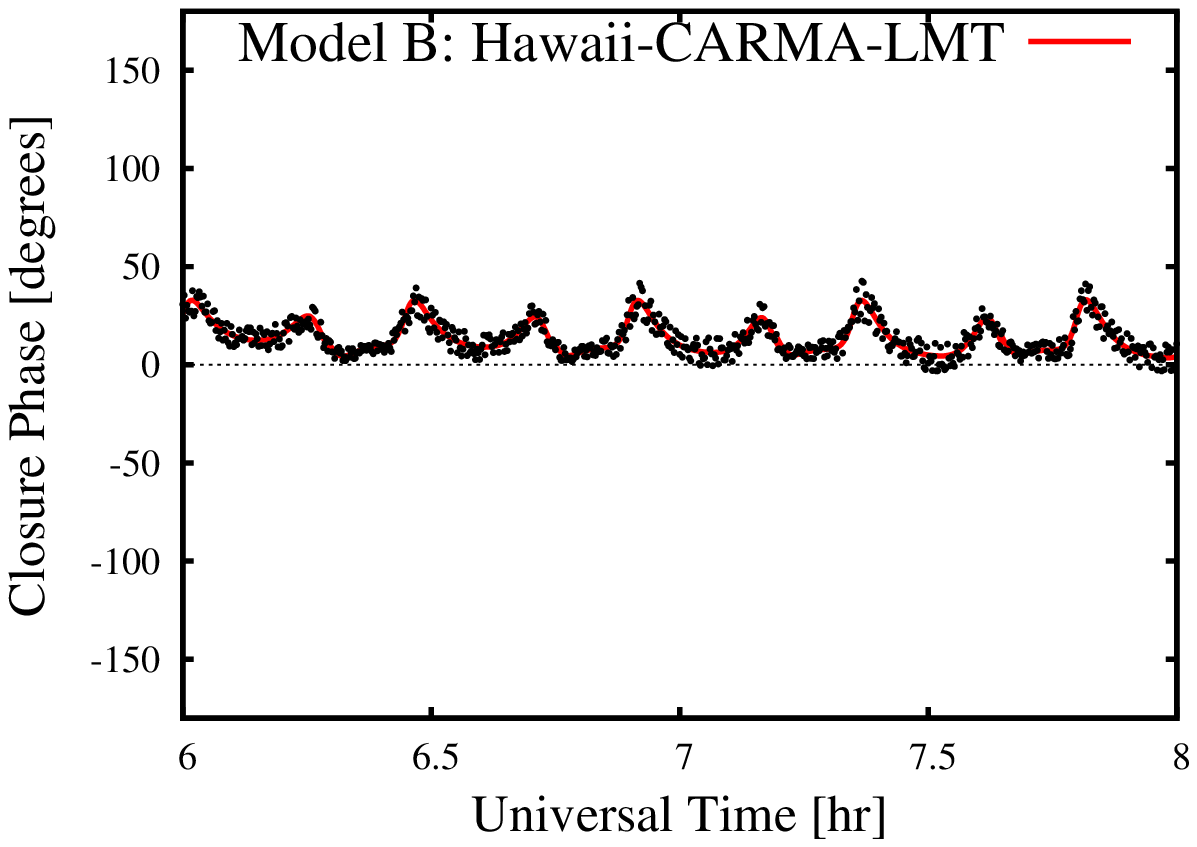}
\includegraphics[width=0.24\textwidth]{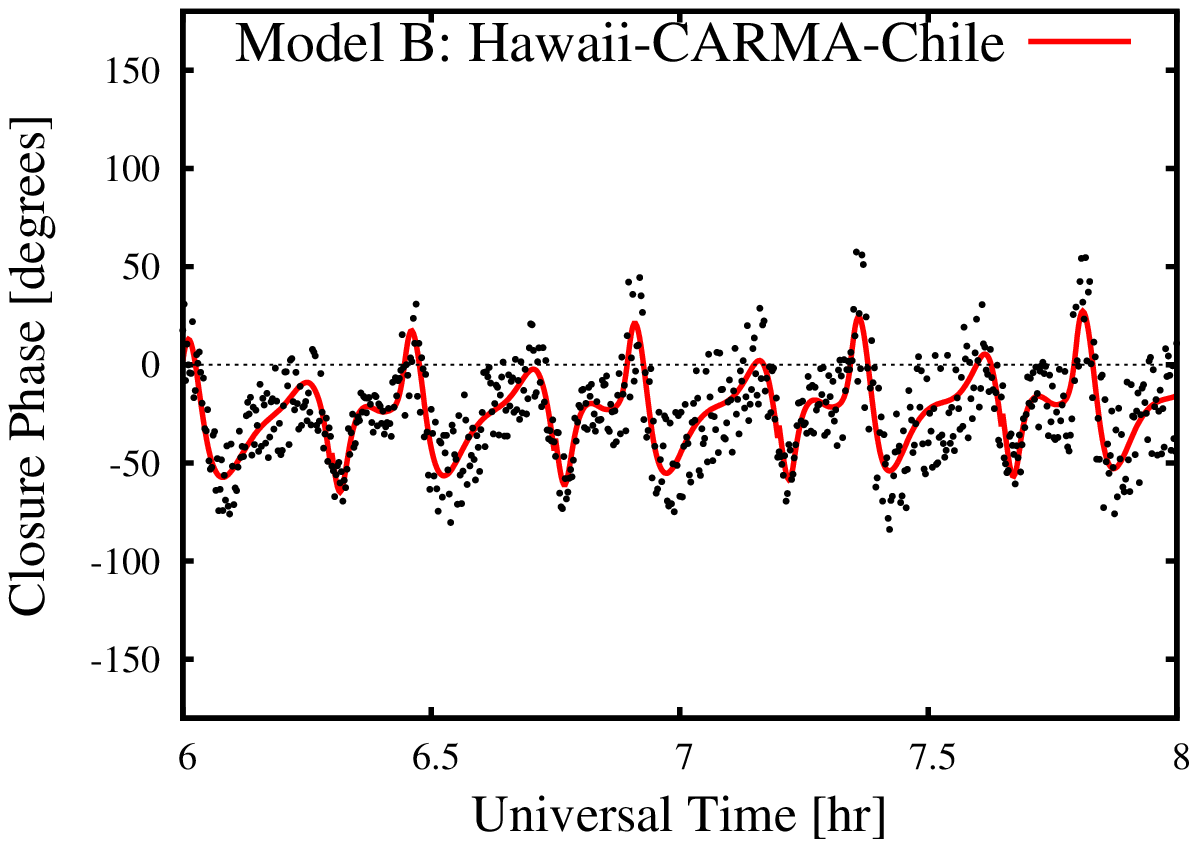}
\includegraphics[width=0.24\textwidth]{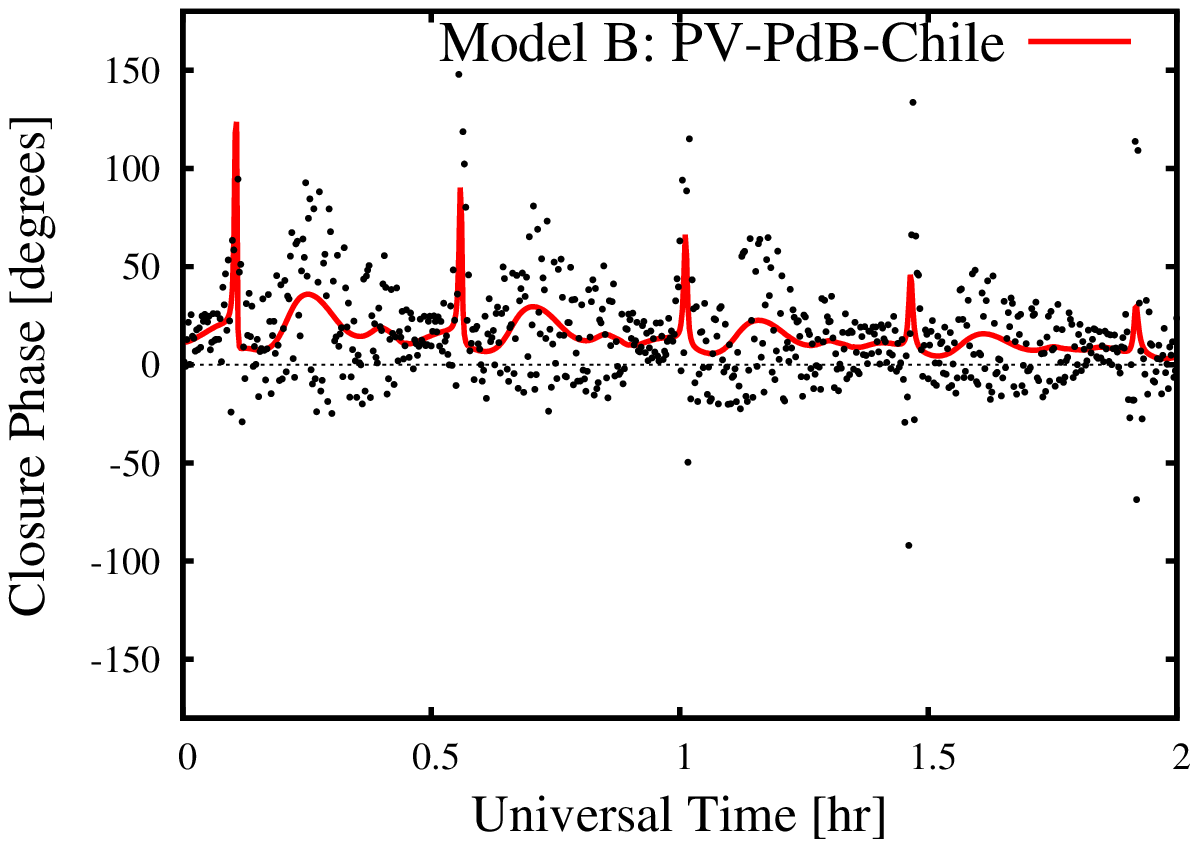}\\
\includegraphics[width=0.24\textwidth]{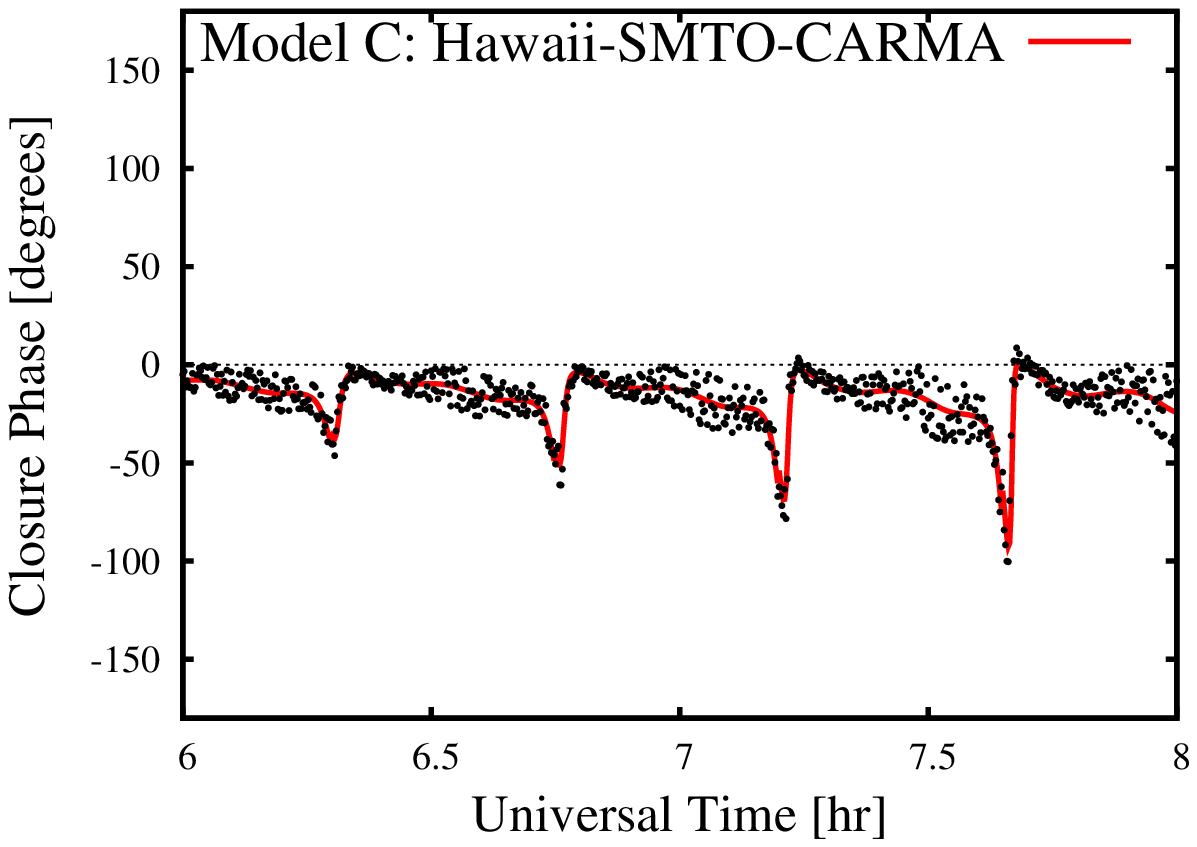}
\includegraphics[width=0.24\textwidth]{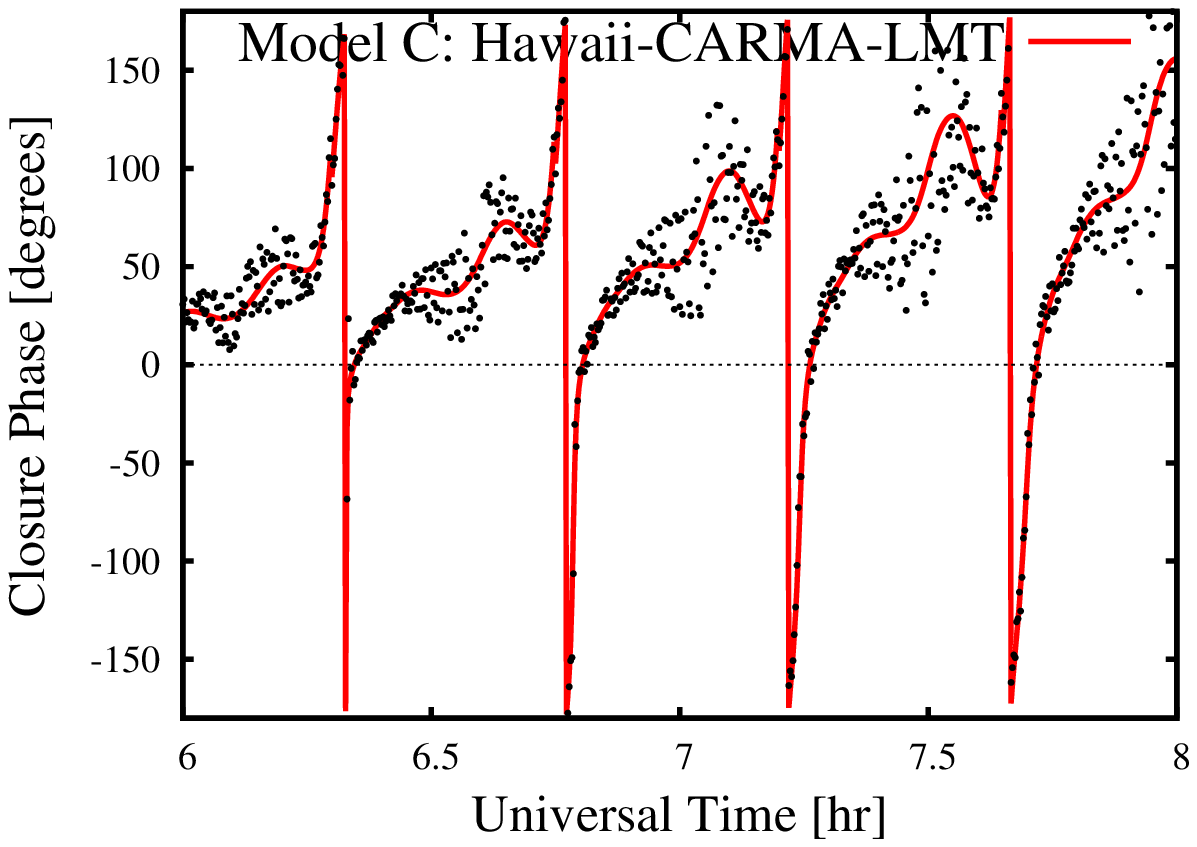}
\includegraphics[width=0.24\textwidth]{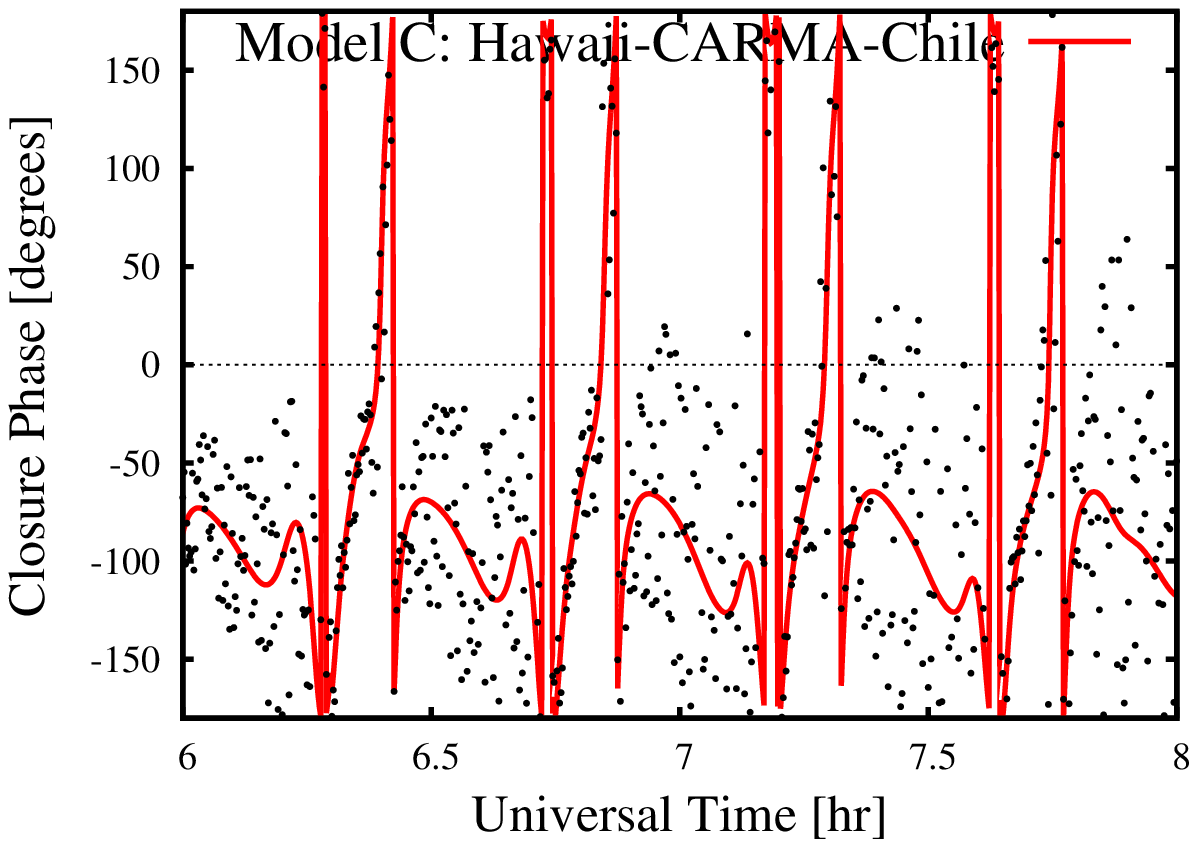}
\includegraphics[width=0.24\textwidth]{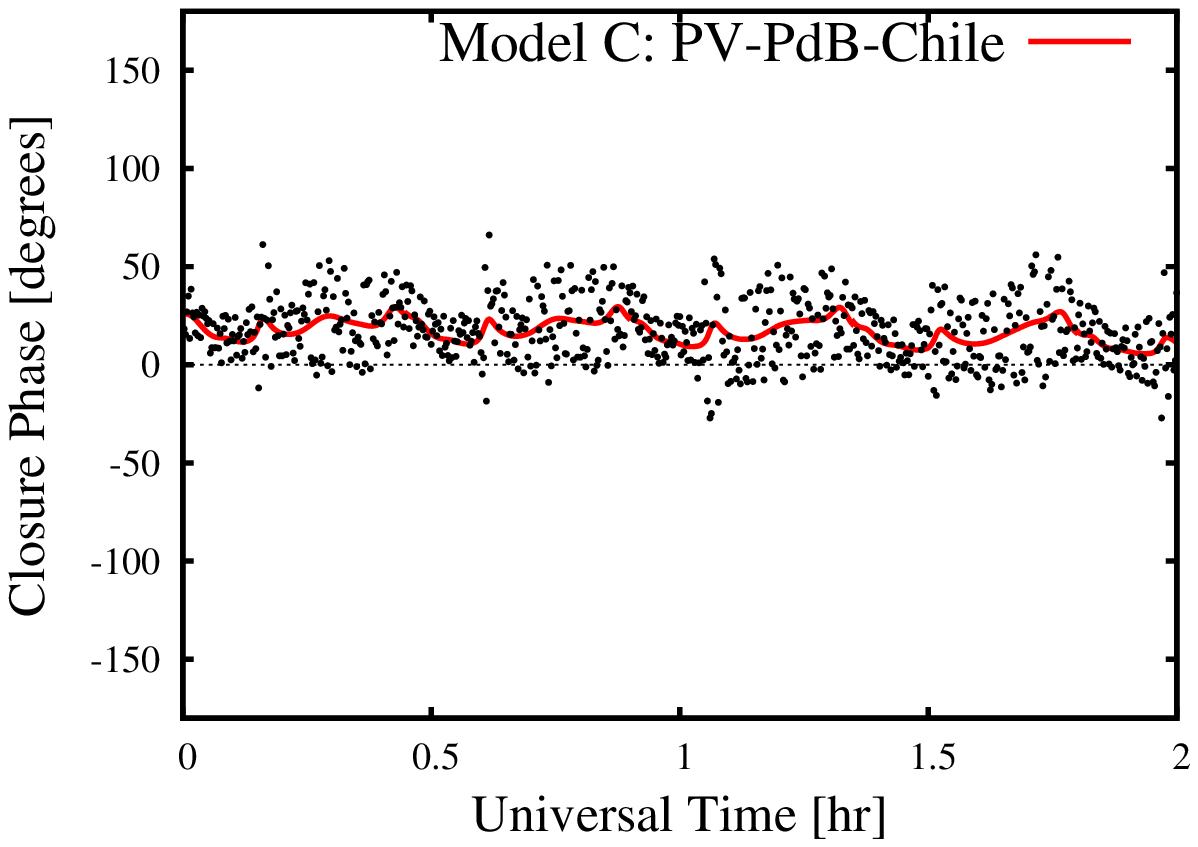}\\
\includegraphics[width=0.24\textwidth]{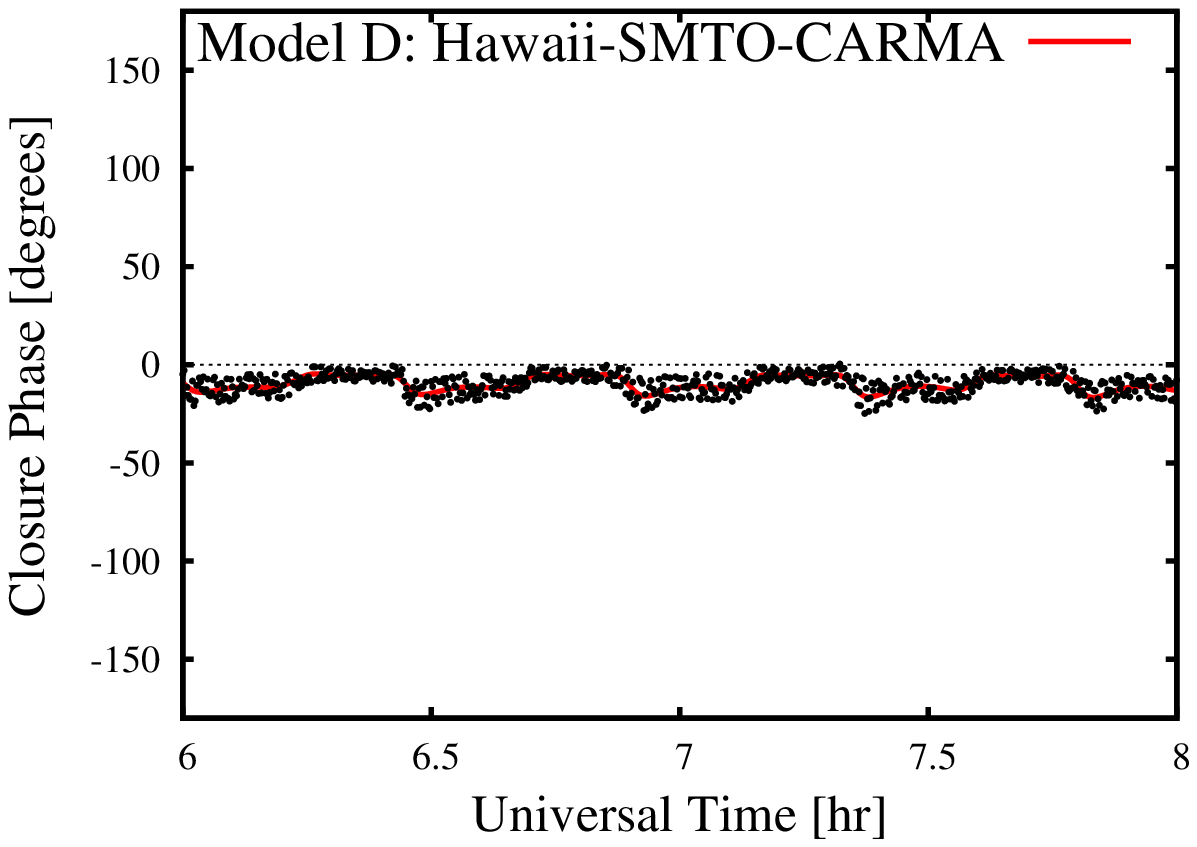}
\includegraphics[width=0.24\textwidth]{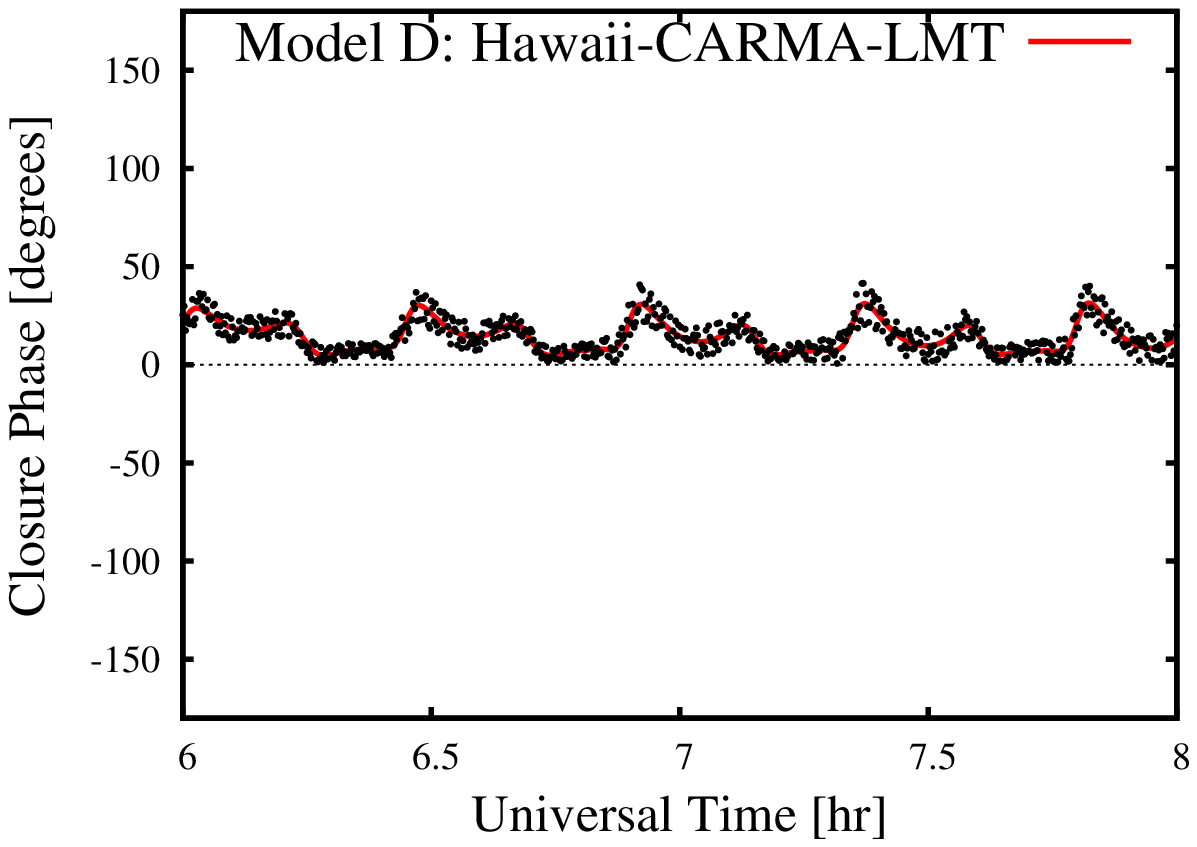}
\includegraphics[width=0.24\textwidth]{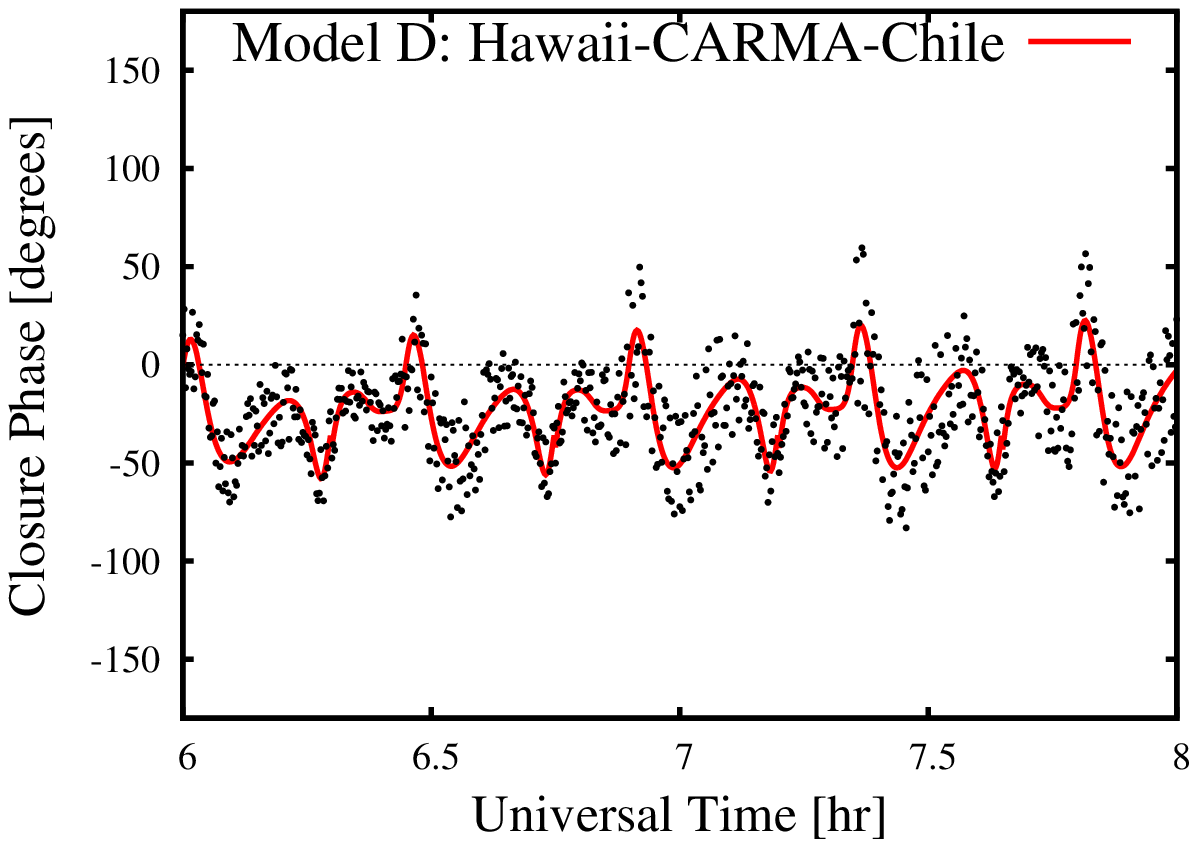}
\includegraphics[width=0.24\textwidth]{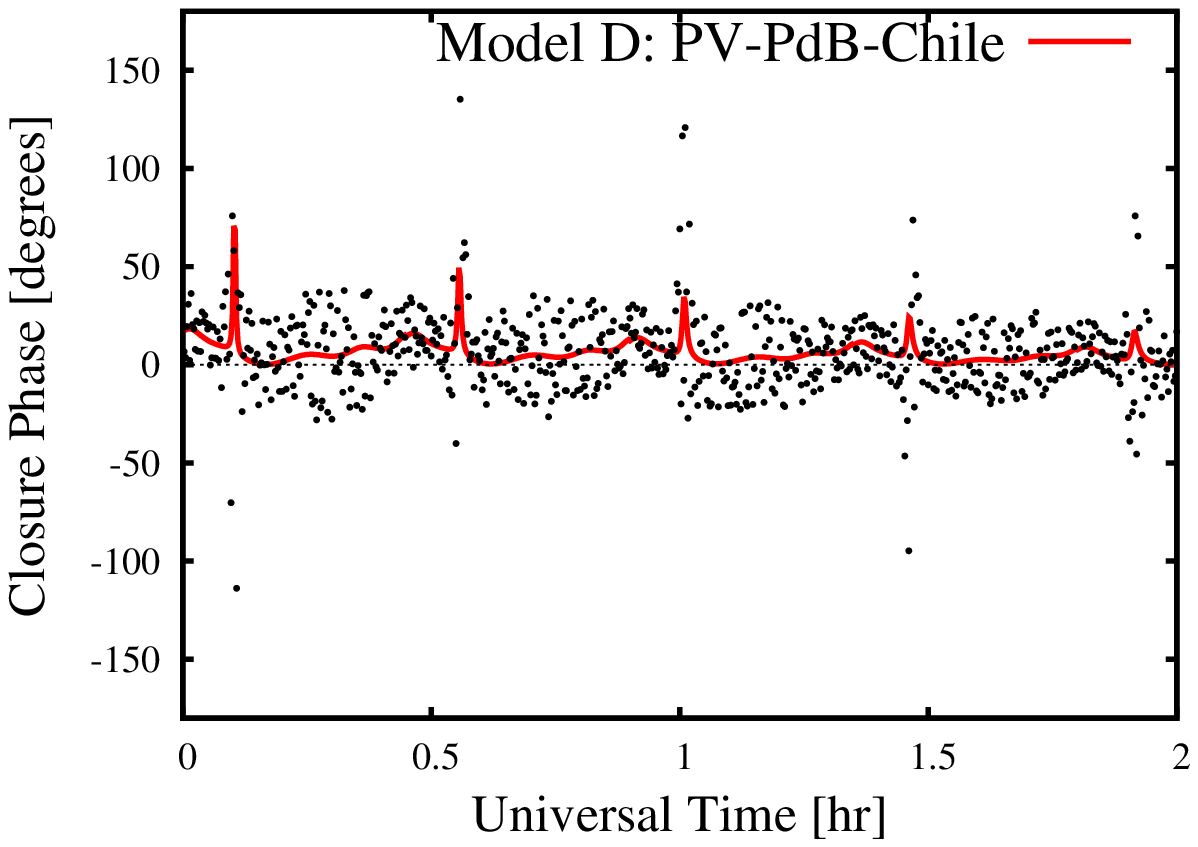}
\caption{Closure phases at $\lambda=$1.3mm on various triangles of VLBI
  stations. Different models of the orbiting spot are presented (from top to
  bottom) with parameters given in Table~\ref{tab:model_params}. Each point
  represents an integration time of 10s, and the same 2h period is shown
  except for the triangle including the European stations. The solid red line
  shows the closure phase without noise. Chile refers to the APEX telescope and Hawaii refers to 3 Hawaiian facilities phased together as described in Table~\ref{tab:SEFD}.}\label{fig:closure_spot}
\end{figure*}

Figure~\ref{fig:closure_spot} shows the predicted closure phase evolution (solid
red line) for
the four orbiting spot models summarized in Table~\ref{tab:model_params}. Each
row from top to bottom shows models from A to D, and each column shows a given
model for the following set of triangles of VLBI stations, from left to right:
Hawaii-SMTO-CARMA, Hawaii-CARMA-LMT, Hawaii-CARMA-Chile, and IRAM PV-IRAM PdB-Chile. 
These models and specific triangles of baselines were chosen to compare our results with those presented
in Fig.~5 by \citet{doeleman:2009}.

All the models display periodicity on the closure phases due 
to the short (27 minutes) orbital period of
the spot around the SMBH. However, we also observe a secular trend in the
closure phase evolution due to the Earth's rotation. At first glance, the most
noticeable feature of the panels in Fig.~\ref{fig:closure_spot} is that the
closure phase prediction depends strongly on the parameters chosen for a given
model. Small triangles of baselines are expected to yield lower values of
the closure phase than large triangles of baselines. This is simply a consequence
of the scale of the features in the brightness distribution on the sky and the
angular resolution that goes with the projected baseline length. Since
interpreting closure phases is highly non-trivial, we can
only safely conclude at this point that closure phases can be used to distinguish between
models with different position and inclination angles.

The observational noise is plotted in Fig.~\ref{fig:closure_spot} as black
points. Here we assume $B=4$ GHz, $T=10$ s, and $\tau=20$ s to calculate 
$\sigma_{\Phi_{ijk}}$ and generate Gaussian random noise. 
The closure phases that include the Chile station assume the SEFD of APEX, hence the
noise is slightly higher than in other triangles. 

Our results (curve shape, phase signs, and evolution of the closure phases), 
although not identical, are roughly 
consistent with those presented in Fig.~5 in \citet{doeleman:2009}.
We conclude that our tools for calculating visibilities and closure phases 
are hereby found to be valid, and therefore we proceed to explore the brightness
distribution functions that are produced by more physically driven scenarios.

\subsection{Theoretical closure phase evolution for the disk and jet models}

\begin{figure}
\centering
\includegraphics[width=0.75\columnwidth]{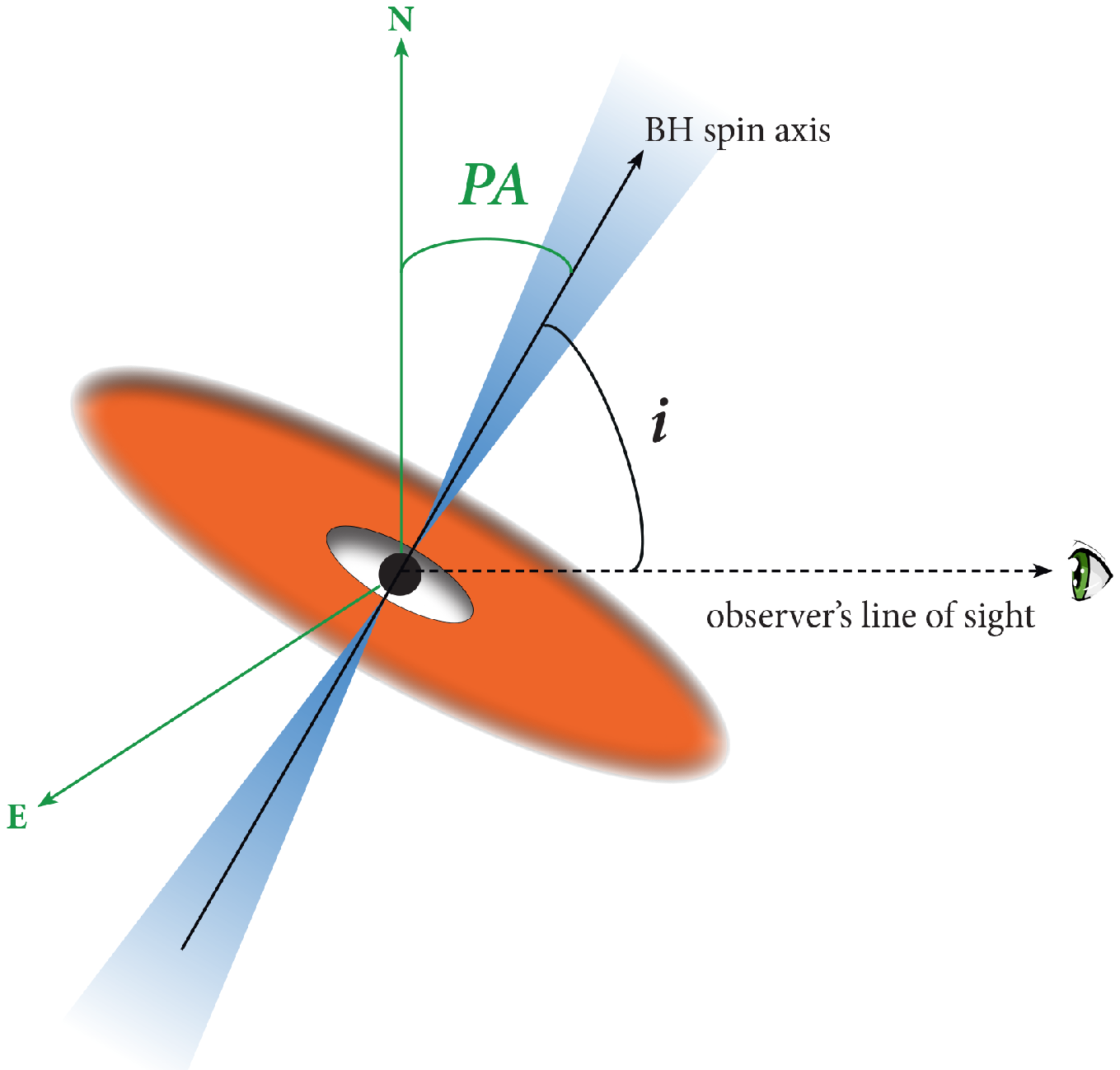}
\caption{Geometry of our models in terms of observer's inclination angle ($i$) and BH spin axis position angle ($PA$).}
\label{fig:BH_geometry}
\end{figure}

\begin{figure*}
\centering
\includegraphics[width=0.25\textwidth,angle=-90]{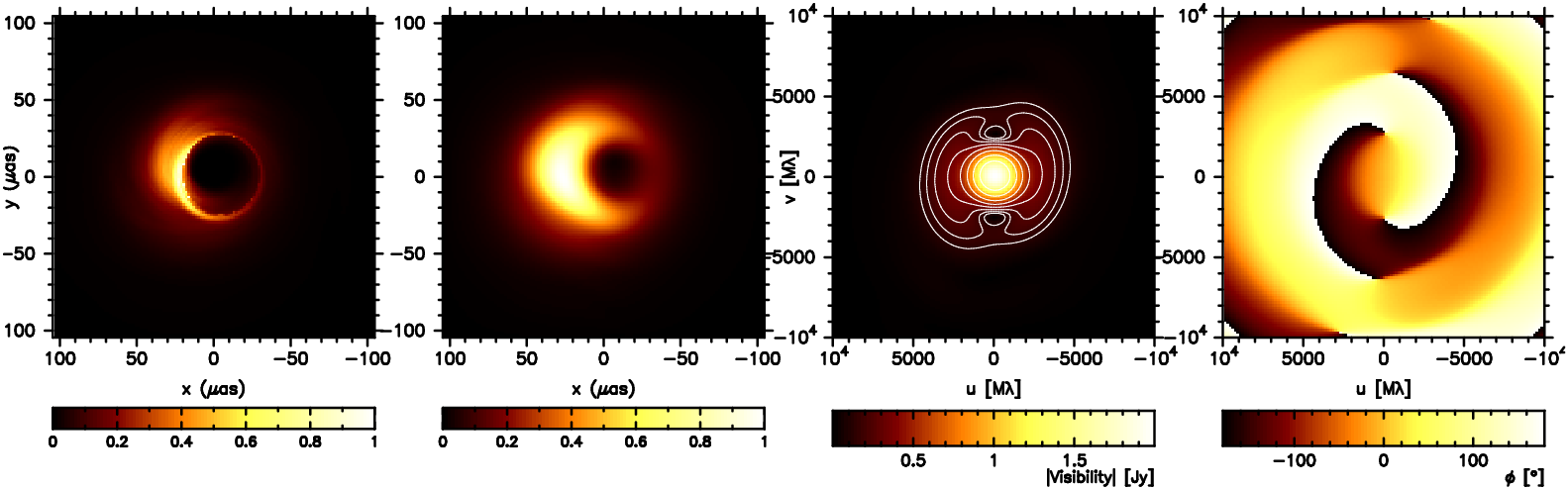}\\
\includegraphics[width=0.25\textwidth,angle=-90]{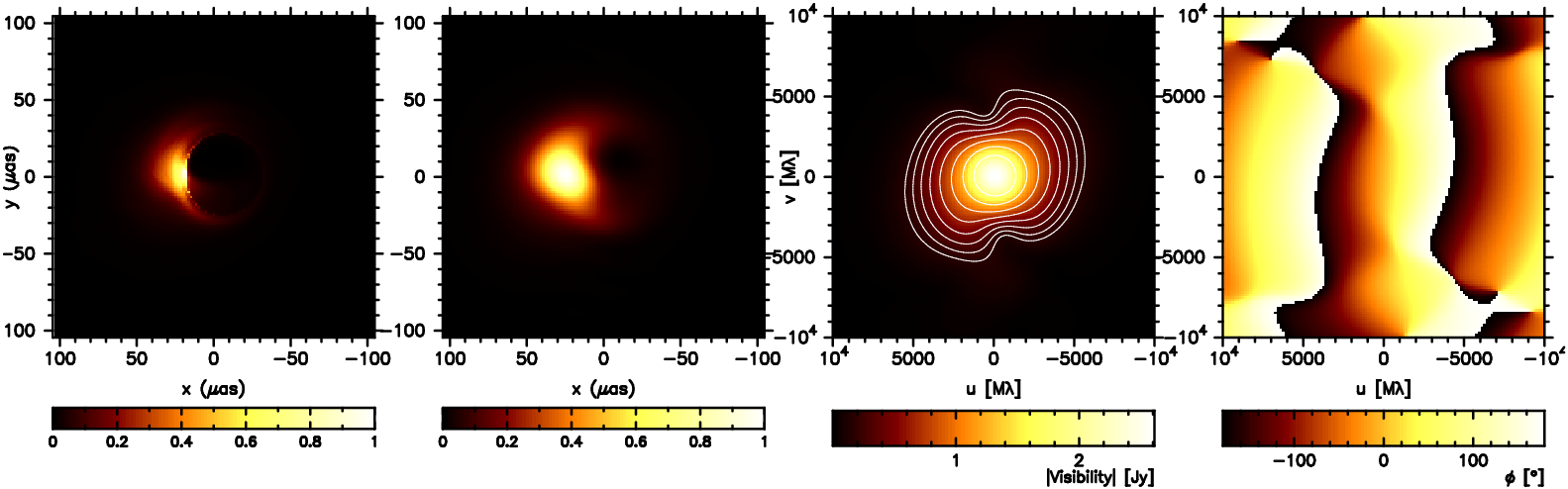}\\
\includegraphics[width=0.25\textwidth,angle=-90]{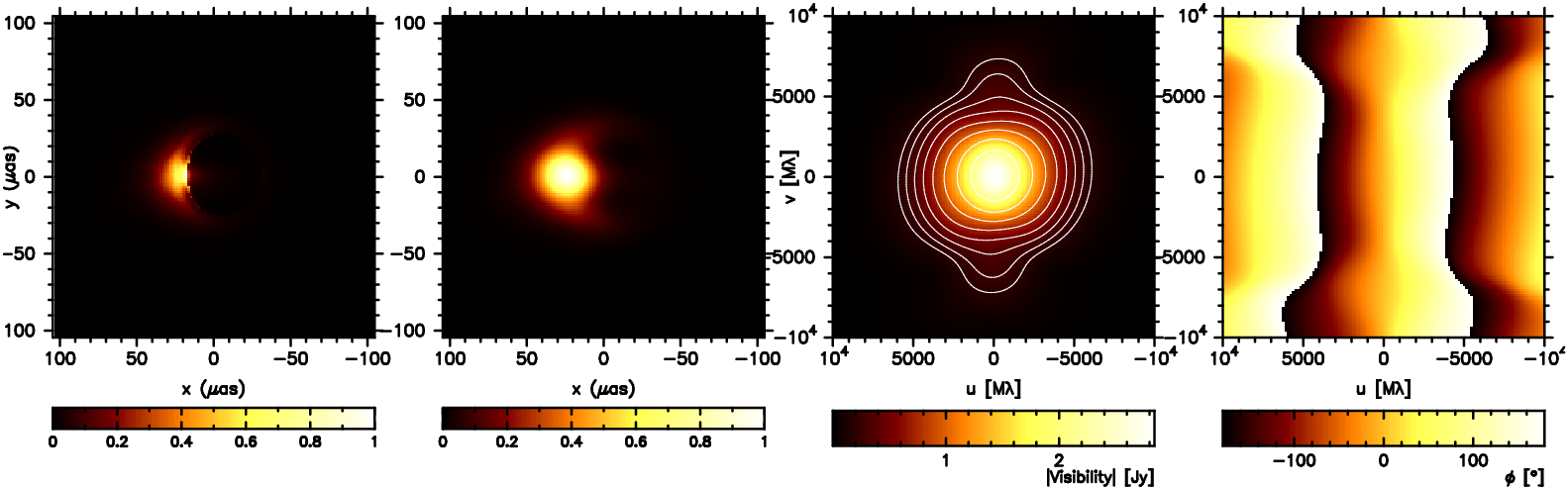}\\
\includegraphics[width=0.25\textwidth,angle=-90]{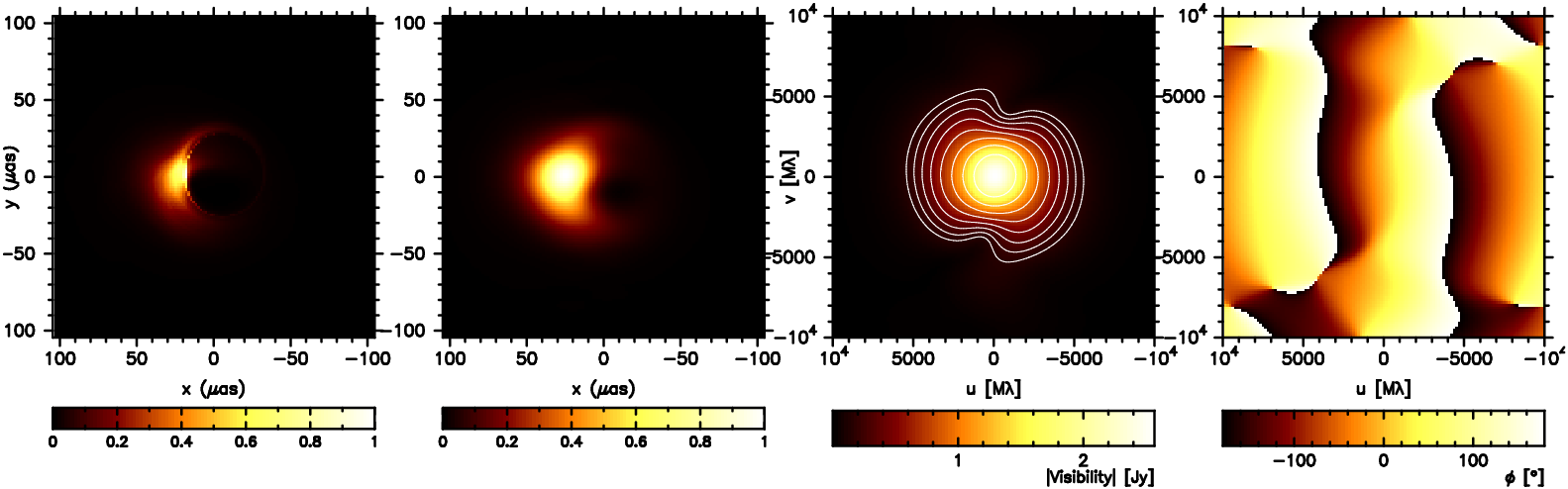}\\
\includegraphics[width=0.25\textwidth,angle=-90]{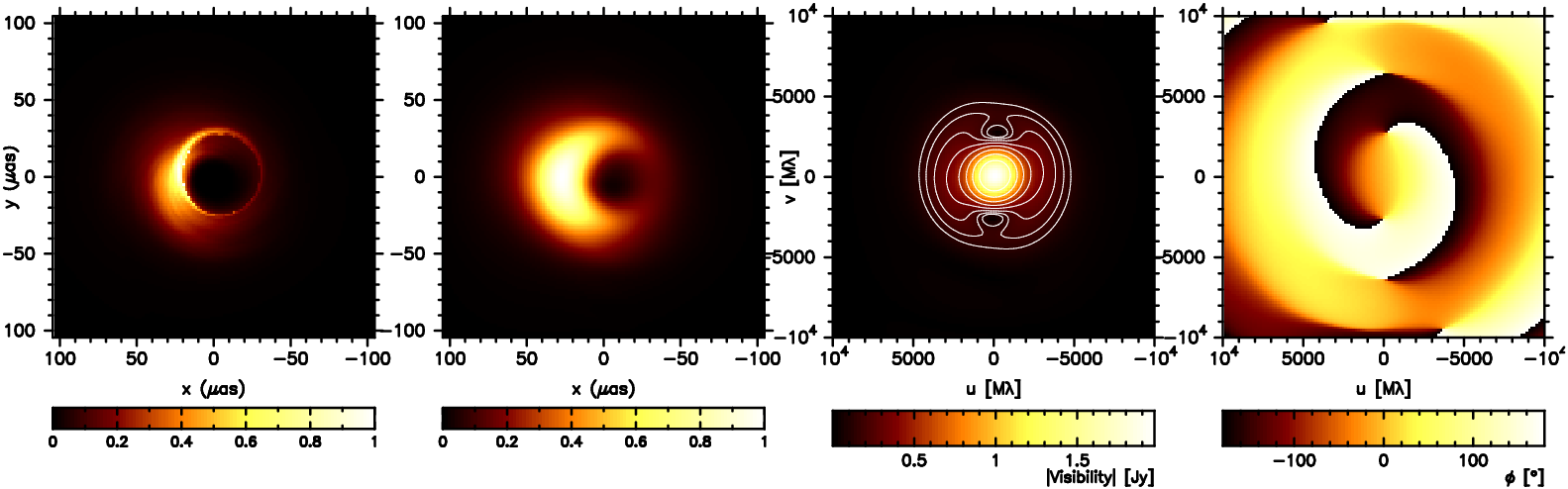}
\caption{Images of the disk model at $\lambda=1.3$mm (here BH spin $PA=0\degr$). Rows
  show inclinations of 30$\degr$, 60$\degr$, 90$\degr$, 120$\degr$, and
  150$\degr$ from top to bottom. Left to right panels show an image of the disk
  model, that image convolved with the scattering screen, the visibility amplitude,  and the visibility phase of the
  scatter-broadened images. The color intensity for the panels in the first two
  columns indicates the intensity of radiation, which has been normalized to unity. In the
  third column the color intensity indicates the amplitude of the visibility
  in Jy and contours are spaced by a factor of $\sqrt{2}$. The last column shows the corresponding map of the visibility phase. The range of uv values for Cols. 3 and 4 is the same.}\label{fig:disk_model}
\end{figure*}

Our disk model is shown in Fig.~\ref{fig:disk_model}. The panels, from left to
right, show the intrinsic image of the disk model, its convolution with the scattering screen, the amplitude of the visibility function (overplotted with
contours), and lastly the phase of
the visibility function. From top to bottom, the inclinations of the disk
range from $i$=30$\degr$ to $i$=150$\degr$, while the BH spin position angle ($PA$) is kept at
0$\degr$ (BH spin is pointing north). In our models, $i$ is the angle between the observer's line of sight and the BH spin axis, with $i$=0$\degr$ being face-on and $i$=90$\degr$ being edge-on. $PA$ is the position angle of the BH spin axis on the sky, where $PA$ is positive in the direction west of north (see Fig.~\ref{fig:BH_geometry}).

In Fig.~\ref{fig:disk_model} the disk model emission is more extended 
than in the RIAF images shown in Fig.~\ref{fig:spot_model}. 
The disk appearance is dominated by the gravitational lensing and relativistic
  Doppler effects. The ring shape is 
evident in the top and bottom panels with inclinations that are more face-on
($i$=30$\degr$ and 150$\degr$). However, at $i$=90$\degr$ the disk appears 
  as a Gaussian-like spot rather than a crescent.
Another important feature to note is that the emission is dominated
by the left side of the disk. This is because the orbiting plasma in the
disk becomes Doppler boosted as it approaches the observer. 
The panels depicting the visibility function, third column, indicate that the black hole shadow is only clearly discernible as two minima in cases when the disk is
  observed face-on. The last column represents the phase of the complex
visibility for the different inclinations.

\begin{figure*}
\centering
\includegraphics[width=0.25\textwidth,angle=-90]{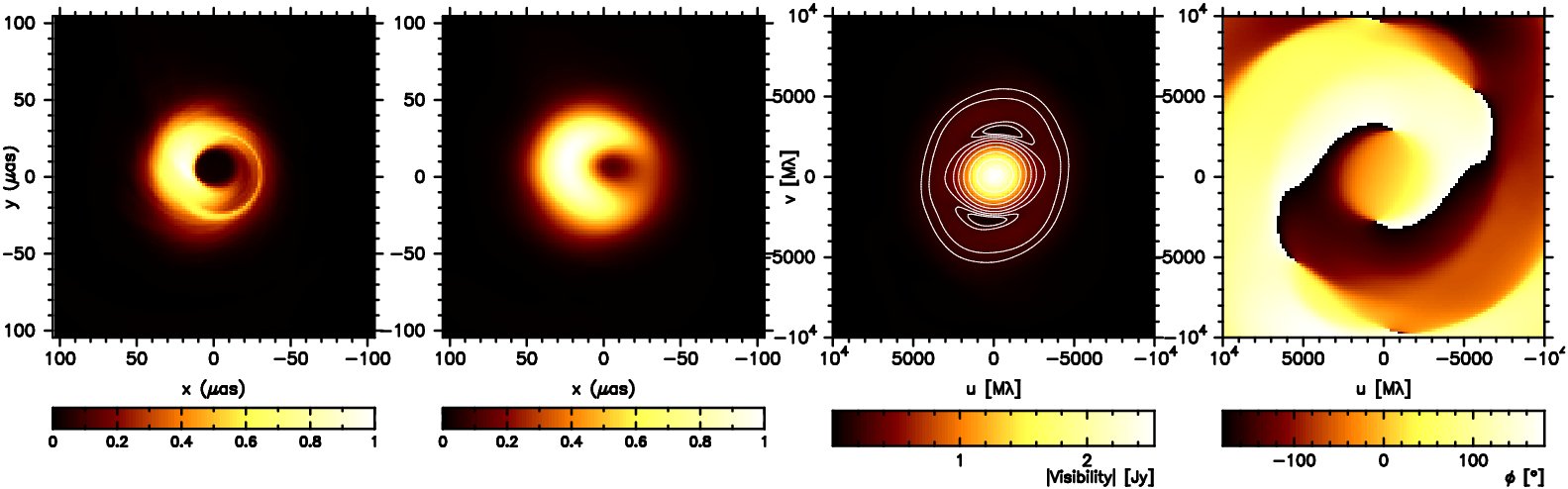}\\
\includegraphics[width=0.25\textwidth,angle=-90]{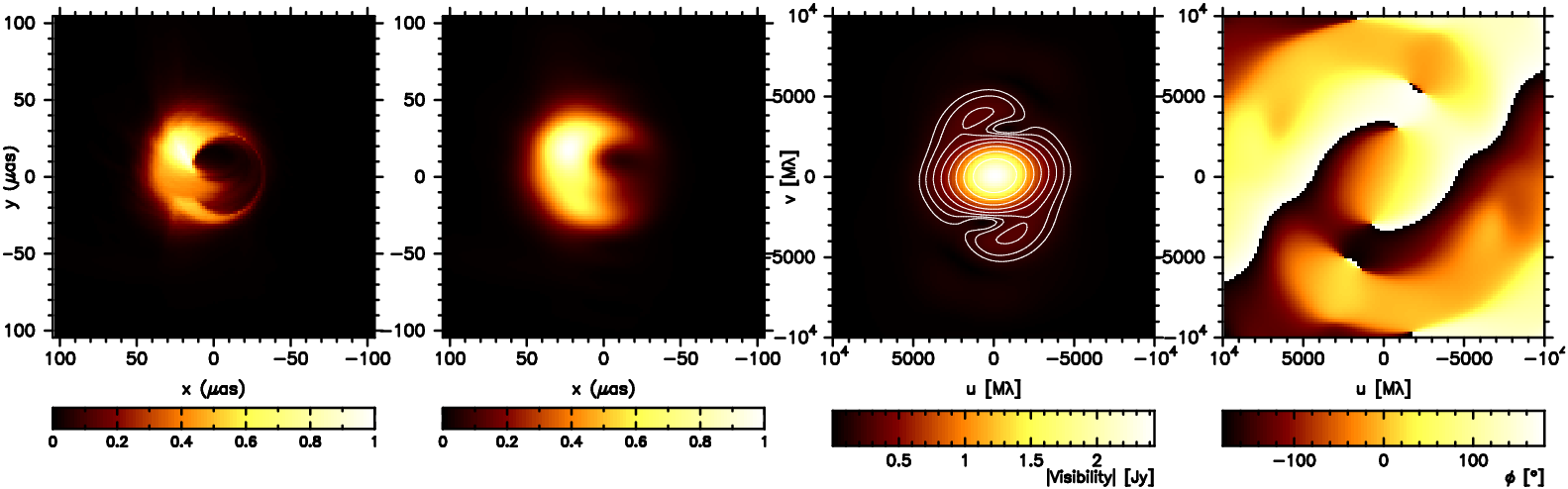}\\
\includegraphics[width=0.25\textwidth,angle=-90]{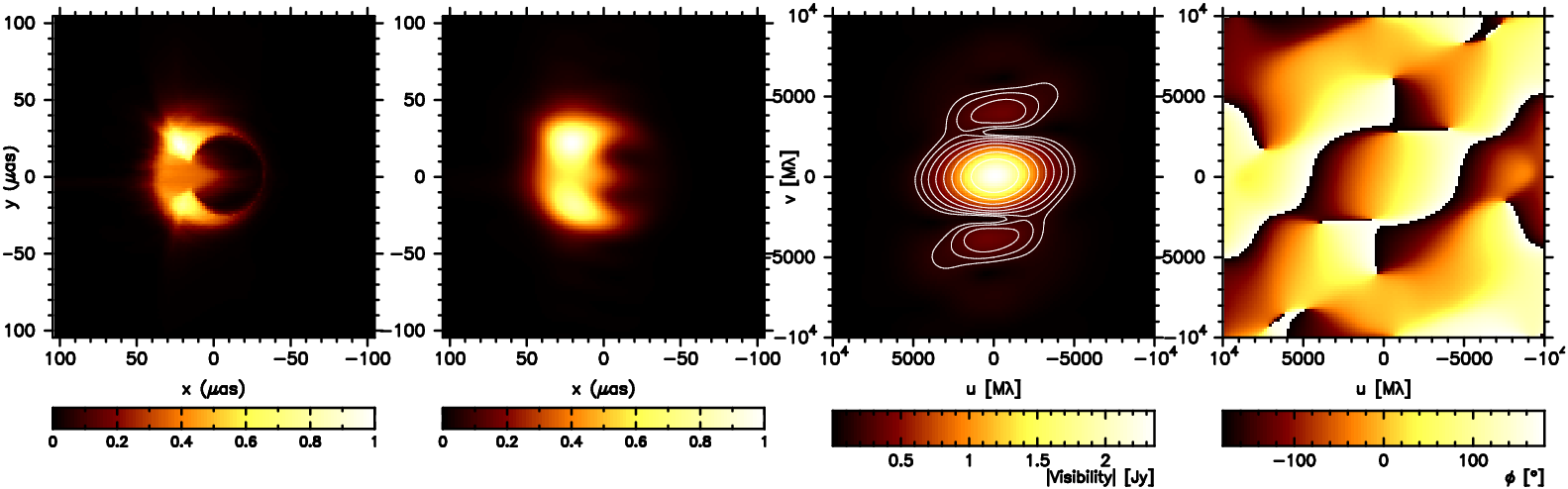}\\
\includegraphics[width=0.25\textwidth,angle=-90]{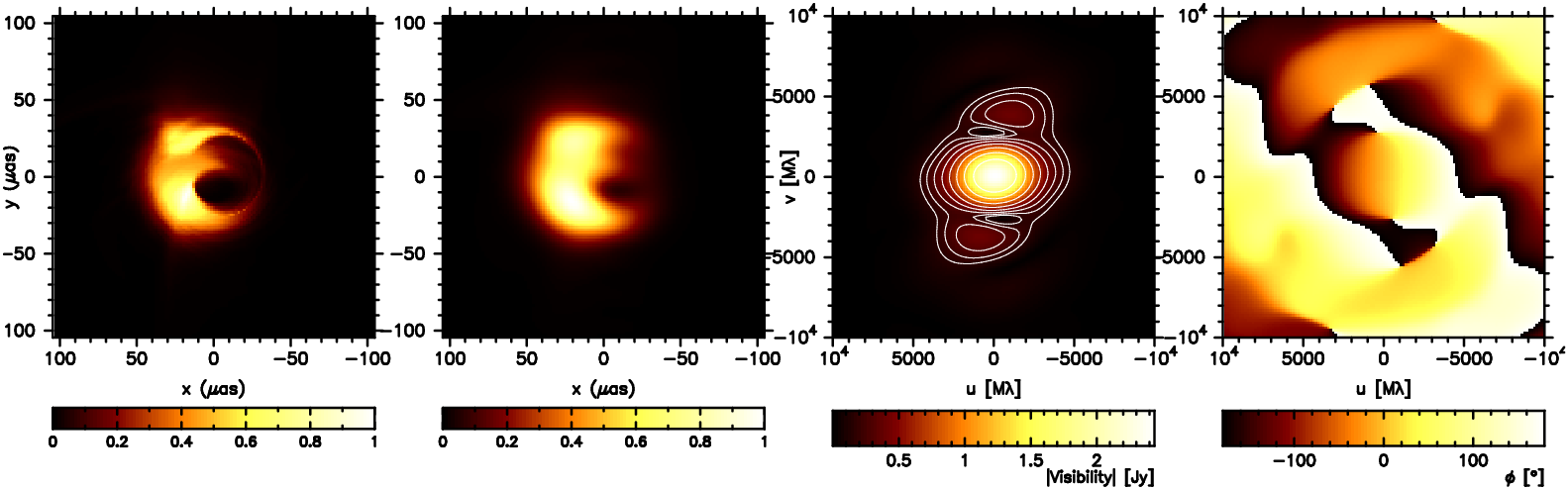}\\
\includegraphics[width=0.25\textwidth,angle=-90]{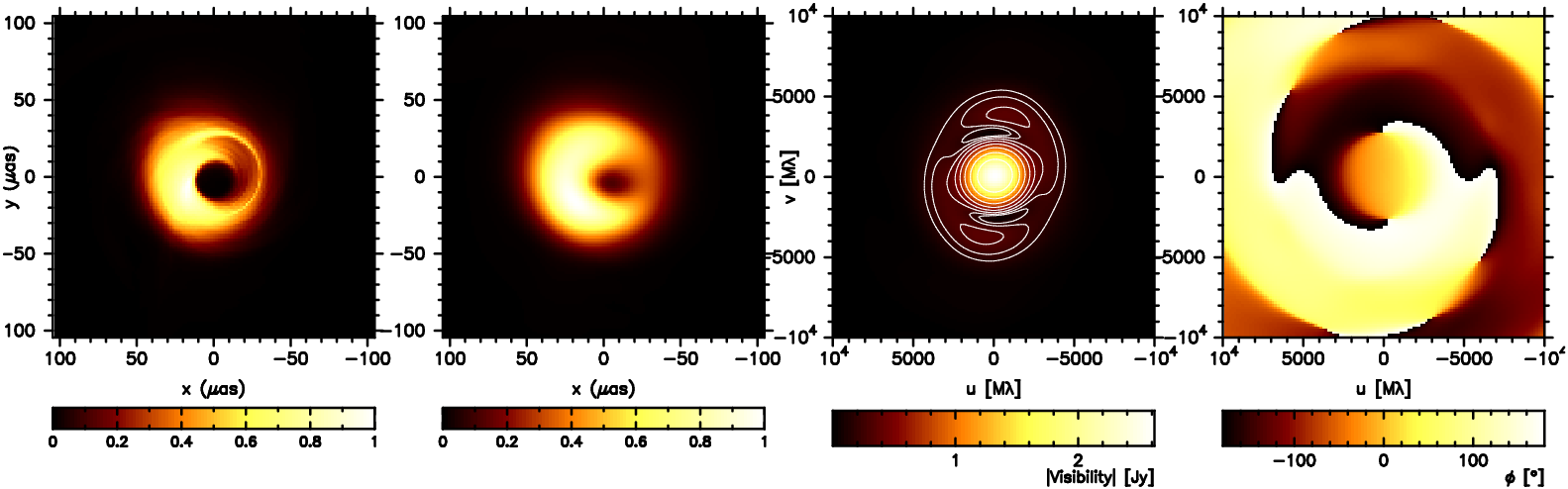}
\caption{Images of the jet model at $\lambda=1.3$mm (here BH spin $PA=0\degr$). Rows
  show inclinations of 30$\degr$, 60$\degr$, 90$\degr$, 120$\degr$, and
  150$\degr$ from top to bottom. Left to right panels show an
image of the jet
  model, that image convolved with the scattering screen, the visibility amplitude,
  and the visibility phase of the
  scatter-broadened images. The color intensity for the panels in the first two
  columns indicates the intensity of radiation, which has been normalized to unity. In the
  third column the color intensity indicates the amplitude of the visibility
  in Jy and contours are spaced by a factor of $\sqrt{2}$. The last column shows the corresponding map of the visibility phase. The range of uv values for Cols. 3 and 4 is the same.}\label{fig:jet_model}
\end{figure*}

The jet model is shown in Fig.~\ref{fig:jet_model} (panels are the same as in
Fig.~\ref{fig:disk_model}). 
Compared to the disk model, the image of a jet near the SMBH horizon
is not well represented by a Gaussian or by a crescent, it has a more
complicated structure. In the jet images, most of the emission is produced by the jet component.
The disk component is weaker, but still visible as a ring (at $i$=30$\degr$
and $i$=150$\degr$), or as a ring plus a tongue-like feature 
that sweeps across the near side of the black hole ($i$=60$\degr$--120$\degr$). 
The jet component is best visible at $i$=90$\degr$ as
two spots separated by a dimmer disk tongue-like feature. These two spots are
the footprints of the large-scale jet.

In general, both the intrinsic and the scatter-broadened images 
of the jet model are more extended compared to the disk images.
One contributing factor is that the accretion rate value for the jet
model is approximately $\dot{M}\approx$ a few $\times 10^{-8} M_{\odot}/yr$, whereas for the disk
model is around $\dot{M}\approx10^{-9} M_{\odot}/yr$. These values are used for the
normalization of the flux at 1.3 mm. Consequently, in our jet scenario
  the black hole shadow is clearly visible
at all viewing angles. In the visibility amplitude panels
the two minima denoting the shadow are detectable regardless of the
inclination angle. Again, the last column represents changes in the phase of the visibility
function with inclination. For $i$=60$\degr$--120$\degr$, the jet model phase maps are significantly
different from the corresponding maps computed based on the disk images.

\begin{figure*}
\centering
\includegraphics[width=0.95\textwidth]{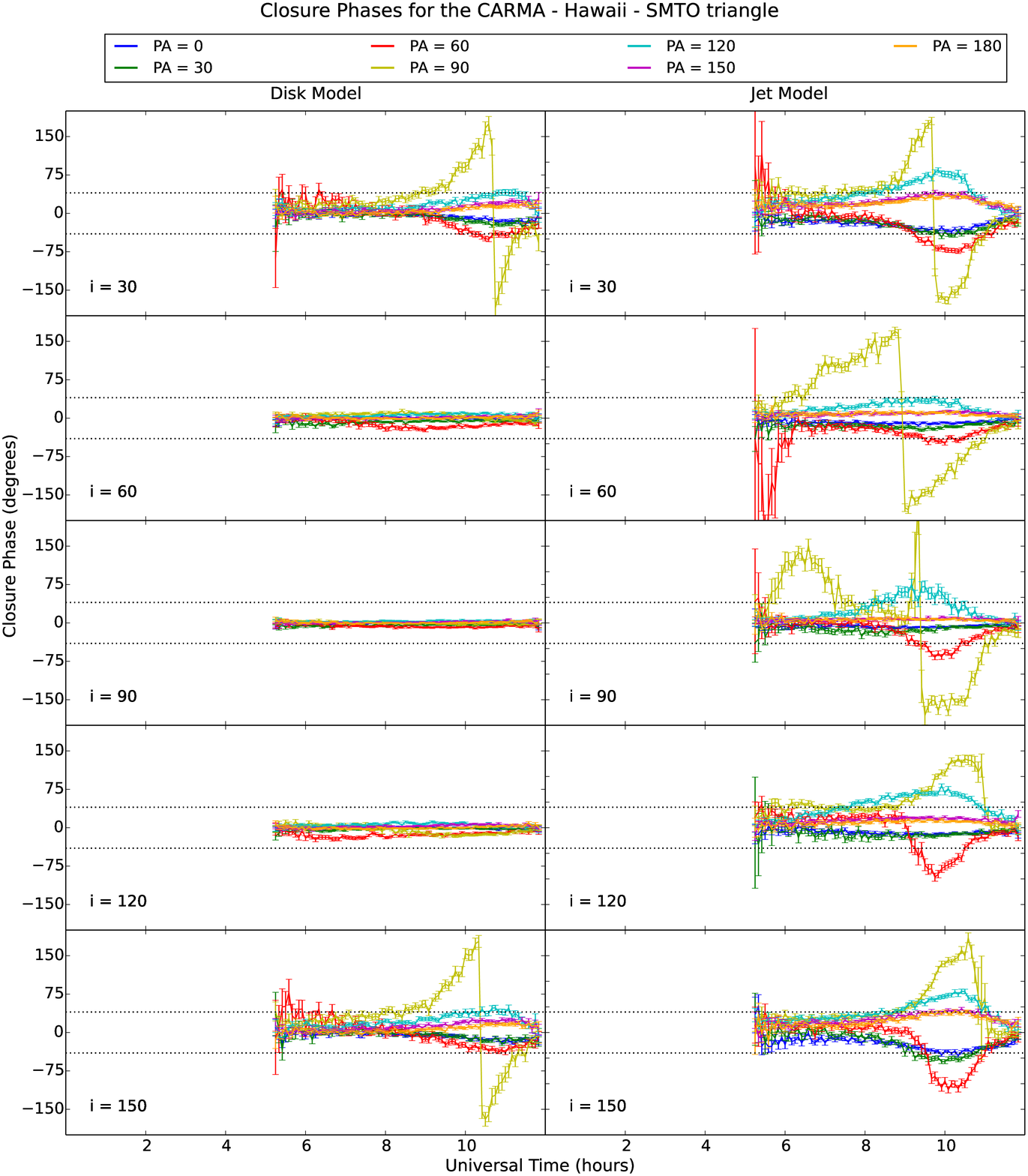}
\caption{Plots of closure phase values at $\lambda=$1.3mm for the disk model and jet
  model for the triangle formed by the CARMA-Hawaii-SMTO baselines. Rows show
  inclination angles between the observer's line of sight and the BH spin axis with values of $i$=30$\degr$, 60$\degr$, 90$\degr$, 120$\degr$, and
  150$\degr$ (from top to bottom) with the left column displaying the disk
  model and the right column the jet model. Solid colored lines represent the
  simulated measurements at different BH spin position angles of $PA$=
  0$\degr$, 30$\degr$, 60$\degr$, 90$\degr$, 120$\degr$, 150$\degr$,
and   180$\degr$ west of north (the images shown in the second column of Figs.~\ref{fig:disk_model} and~\ref{fig:jet_model} are rotated in the clock-wise
  direction). The horizontal dotted lines represent values of $\pm$ 40$\degr$
  closure phase as measured by \citet{fish:2011} over a 3.5 hour interval. The gap between UT=0--5
  hours is due to the lack of baselines visible from Sgr A* during this period
  of time for this triangle.}\label{fig:jetdisk_CARMA_Hawaii_SMT0}
\end{figure*}

The closure phases at 1.3 mm for the CARMA-Hawaii-SMTO triangle are shown in
Fig.~\ref{fig:jetdisk_CARMA_Hawaii_SMT0} for both the disk and jet
models. \textit{\textup{This is the main new result delivered by this work}}.  In
Fig.~\ref{fig:jetdisk_CARMA_Hawaii_SMT0} the colored lines represent the
predicted closure phases for different BH spin position angles ($PA$ is
  positive in the direction west of north, i.e., we rotated the images shown in the second column of Figs.~\ref{fig:disk_model} and~\ref{fig:jet_model} in the
  clock-wise direction, which is opposite of the normal convention) and the panels from top to bottom show
different inclinations. We present closure phases for $PA$s ranging from 0$\degr$ to
180$\degr$. Position angles offset by 180$\degr$ yield mirror results. We
plot a $T$=5 minute average of the closure phase (average of thirty 10-second
scans) with the corresponding error bars, spanning a measurement period of
about seven hours.  In Fig.~\ref{fig:jetdisk_CARMA_Hawaii_SMT0} we have
included dotted lines to indicate a closure phase value of $\pm$ 40$\degr$ as was measured over a 3.5 hour interval by \citet{fish:2011}.

First, for both models the closure phases are the largest 
at face-on viewing angles, as expected for the ring-like structures.
Second, the most obvious difference between the closure phases of the
disk model and the jet model is that for the jet model they tend to have
higher values at viewing angles $i$=60--120$\degr$. This is expected since closure phases
give us information about the amount of asymmetric flux the source has, and we
can see in the images of the jet model (see Fig.~\ref{fig:jet_model}) that
their structure is more asymmetric than that of the disk model.

Based on the $\pm$ 40$\degr$ closure phase measurements by \citet{fish:2011}
(dotted lines in Fig.~\ref{fig:jetdisk_CARMA_Hawaii_SMT0}), models
can already be constrained.  For example, the disk model, at inclinations of
  $i$=60$\degr$ to 120$\degr$ cannot be ruled out for any PAs since all have
  closure phase values that are close to zero. For more face-on inclinations
  (30$\degr$ and 150$\degr$), a disk model with BH spin axis position angle of
  $PA$=60$\degr$, 90$\degr$, or 120$\degr$ can be discarded since the predicted
  closure phases fall outside the measured values. For all inclinations of the jet model, position angles of $PA$=60$\degr$, 90$\degr$, and 120$\degr$ are clearly inconsistent with the preliminary observations. Other
  orientations require further inspection.

\begin{figure*}
\centering
\includegraphics[width=0.95\textwidth]{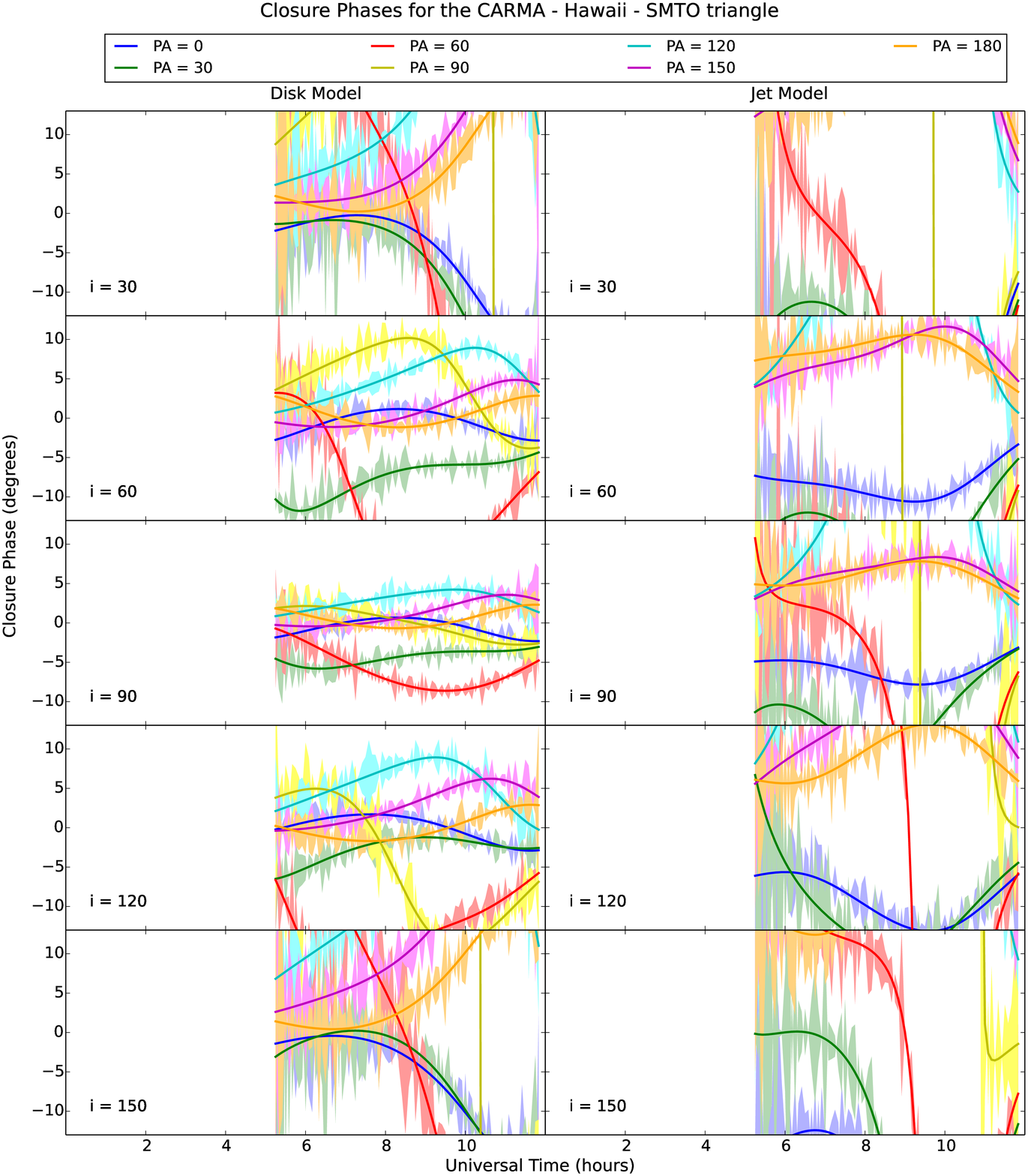}
\caption{Same as Fig.~\ref{fig:jetdisk_CARMA_Hawaii_SMT0}, but zooming-in into
  the area of interest. The y-axis range has been set to $\pm$ 13$\degr$, this
  limit represents the most likely values for closure phase as shown in the
  models by \citet{broderick:2011_closure}. Solid lines are closure phases
  without noise, shaded areas represent the noise for the corresponding
  curve.}\label{fig:jetdisk_CARMA_Hawaii_SMT0_ZOOM}
\end{figure*}

Figure~\ref{fig:jetdisk_CARMA_Hawaii_SMT0_ZOOM} depicts the same jet and disk
models, but zooming-in into the area of interest 
($\pm$ 13$\degr$ limits from \citet{broderick:2011_closure} 
are the most likely limits based on accretion flow models consistent with measured visibility
amplitudes). The
difference in the predicted values of closure phases based on the studied
models can also be examined. Jet and disk geometries can reproduce similar closure phases 
for different sets of $i$ and $PA$. Consequently, the black hole spin orientation 
(and possibly the spin value) will be strongly model dependent. 
We conclude that new observations at 1.3 mm (and possibly simultaneous
observations at longer
wavelengths) including other triangles of baselines are necessary to 
constrain the source geometry and the orientation of the black hole spin, for instance.

\section{Discussion}\label{discussion}

We have constructed a number of images of the SMBH shadow for
various surrounding plasma configurations and have predicted the observed
1.3 mm VLBI closure phases. 

Previous work on RIAF models has resulted in
most probable values for the BH spin of $a_*$ $\approx$ 0, inclination of $i$
$\approx$ 68$\degr$ and position angles of $PA$ $\approx$ -52$\degr$ or
+128$\degr$ east of north (\citealt{broderick:2011_models}). Their values of
$PA$ $\approx$ -52$\degr$ or +128$\degr$ would be consistent with our disk
models with inclinations ranging from 60$\degr$ to 120$\degr$, but our jet
model at corresponding $i=$60$\degr$ and PA=60$\degr$ west of north
is inconsistent with the observational limits.
In previous 3D GRMHD models of a disk, 
although quite dynamically different from the RIAF models, similar
favored values of $i$ $\approx$ 50$\degr$ and $PA$ $\approx$ -23$\degr$ east of north 
have been reported for instance by \citet{dexter:2010} based on visibility amplitude analyses. 

As is evident from the jet models (Figs.~\ref{fig:jetdisk_CARMA_Hawaii_SMT0}
and~\ref{fig:jetdisk_CARMA_Hawaii_SMT0_ZOOM}), the closure phases show
more variability and higher values at close to edge-on inclinations than for
the disk models, and more possibilities for orientations are discarded if we
follow the constraint of a modeled closure phase of $\pm$ 13$\degr$ compared
to a measured closure phase of $\pm$ 40$\degr$. Nevertheless, we should not be too
quick to discard the possibility that Sgr~A* has a mildly relativistic
jet. Observational signatures such as a flat-to-inverted radio spectrum
resembling that of an AGN, changes in source size depending on the observing
frequency (\citealt{bower:2004}), time lags between flares at 43 GHz and 22
GHz indicating the presence of relativistic outflows
(\citealt{yusef-zadeh:2006}), and Chandra X-ray observations suggesting an
outflow from the accretion flow (\citealt{wang:2013}) are all clues indicating the presence of a jet. 

In our jet models the preferred orientations
that fall within observational and modeled closure phase limits are
$PA$=0$\degr$, 30$\degr$, 150$\degr$, and 180$\degr$, and preferred
inclinations are more edge-on, so the jet would seem to be pointing close to
the plane of the sky. In terms of the inclination angle, but not necessarily
position angle, our results seem to agree with previous analyses of
7 mm data that favored a highly inclined jet \citep{markoff:2007}. Regarding the
systematic uncertainties in this work, it is important mention that we sample the parameter space sparsely in our models, both the inclination and
position angle change in increments of 30$\degr$. As a result, we do not cover
the whole parameter space. Therefore, it would be possible to have models
that are not represented here and are still consistent with the closure phase
limits described.

Furthermore, all the models investigated here have the axis of the BH angular
momentum aligned with the accretion disk axis. 
Past work has posed the idea that the accretion disk and the black hole angular
momentum axes do not necessarily have to be aligned (tilt of 15$\degr$), and
that it is possible to reproduce the observed time-variable mm and NIR
emission and the crescent shape of Sgr~A* with a tilted model
(\citealt{dexter:2013}). Because we have not yet investigated the dependence
of closure phase on BH spin, magnetic field strength (free parameters), and
disk and black hole spin axes alignment, 
the conclusions presented here might change in the future if a more
comprehensive study of magnetized models and spin models is carried out.

Finally, it is important to note that we have examined emission models based on standard GR, but other studies
have been conducted using a quasi-Kerr metric to predict the image of the accretion
flow (\citealt{johannsen:2010}, \citealt{broderick:2014}).

\section{Conclusions}\label{conclusions}
Disk and jet models of Sgr~A* have been investigated to try to constrain the intrinsic geometry of the source by using closure phase values. We showed that a significant fraction of the disk and jet models can be excluded for a certain combination of the parameters observer’s inclination angle and BH spin axis position angle. However, we cannot yet distinguish between disk and jet models given the range of allowed parameters for both models. Similar analyses of observations and simulations need to be conducted at wavelengths of 0.8 mm, 3.5 mm, and 7 mm to further constrain the emission models for Sgr~A*. At these wavelengths the emission regions probed will be very different to those presented here. At VLBI wavelengths longer than 1.3 mm interstellar scattering broadens the images of Sgr~A* , which tends to dilute the structural information. At VLBI wavelengths much shorter than 1.3 mm observations are not possible because the atmosphere becomes optically thick. Although the disk and jet models have a different appearance because the geometry and emission mechanism are intrinsic properties of Sgr~A*, consistent values for the observer’s inclination angle and BH spin position angle are expected for the aforementioned wavelengths. Hence, analyses of closure phases at 0.8 mm, 3.5 mm, and 7 mm will help to distinguish among the types of emission models. Additional measurements of closure phases in more baseline triangles will help us to distinguish even better among the possible model solutions
because this work was mainly focused on the CARMA-Hawaii-SMTO triangle of stations.

\begin{acknowledgements}
This work is partially supported by the NWO Spinoza Prize awarded to Heino Falcke and by
the ERC Synergy Grant ``BlackHoleCam: Imaging the Event Horizon of Black Holes''
awarded to Heino Falcke, Michael Kramer and Luciano Rezzolla.
\end{acknowledgements}

\bibliographystyle{aa}
\bibliography{local}

\end{document}